\documentclass[aps,prd,twocolumn,showpacs,superscriptaddress]{revtex4}
\usepackage{graphicx}
\usepackage{dcolumn}
\usepackage{bm}
\usepackage{amsmath}
\usepackage{amssymb}

\renewcommand{\arraystretch}{1.1}

\begin{document}

\preprint{\vbox{ 
                 \hbox{Version 12 \today}
                              }}

\title{\quad\\
Observation of an enhancement in $e^+e^- \to \Upsilon(1S)\pi^+ \pi^-$, $\Upsilon(2S)\pi^+ \pi^-$,
and $\Upsilon(3S)\pi^+ \pi^-$ production near $\sqrt{s}=10.89$ GeV}

\begin{abstract}
We measure the production cross sections for $e^+e^- \to
\Upsilon(1S)\pi^+\pi^-$, $\Upsilon(2S)\pi^+\pi^-$, and
$\Upsilon(3S)\pi^+\pi^-$ as a function of $\sqrt{s}$ between 10.83
GeV and 11.02 GeV. The data consists of 8.1 fb$^{-1}$ collected
with the Belle detector at the KEKB $e^+e^-$ collider. We observe
enhanced production in all three final states that does not
agree well with the conventional $\Upsilon(10860)$ lineshape. A
fit using a Breit-Wigner resonance shape yields a peak mass of
$\left[10888.4^{+2.7}_{-2.6} ({\rm stat}) \pm1.2 ({\rm
syst})\right]$ MeV/$c^2$ and a width of $\left[30.7_{-7.0}^{+8.3}
({\rm stat}) \pm3.1 ({\rm syst})\right]$ MeV/$c^2$.
\end{abstract}

\affiliation{Budker Institute of Nuclear Physics, Novosibirsk}
\affiliation{University of Cincinnati, Cincinnati, Ohio 45221}
\affiliation{Department of Physics, Fu Jen Catholic University, Taipei}
\affiliation{Justus-Liebig-Universit\"at Gie\ss{}en, Gie\ss{}en}
\affiliation{The Graduate University for Advanced Studies, Hayama}
\affiliation{Hanyang University, Seoul}
\affiliation{University of Hawaii, Honolulu, Hawaii 96822}
\affiliation{High Energy Accelerator Research Organization (KEK), Tsukuba}
\affiliation{Hiroshima Institute of Technology, Hiroshima}
\affiliation{Institute of High Energy Physics, Chinese Academy of Sciences, Beijing}
\affiliation{Institute of High Energy Physics, Vienna}
\affiliation{Institute of High Energy Physics, Protvino}
\affiliation{Institute for Theoretical and Experimental Physics, Moscow}
\affiliation{J. Stefan Institute, Ljubljana}
\affiliation{Kanagawa University, Yokohama}
\affiliation{Korea University, Seoul}
\affiliation{Kyungpook National University, Taegu}
\affiliation{\'Ecole Polytechnique F\'ed\'erale de Lausanne (EPFL), Lausanne}
\affiliation{Faculty of Mathematics and Physics, University of Ljubljana, Ljubljana}
\affiliation{University of Maribor, Maribor}
\affiliation{Max-Planck-Institut f\"ur Physik, M\"unchen}
\affiliation{University of Melbourne, School of Physics, Victoria 3010}
\affiliation{Nagoya University, Nagoya}
\affiliation{Nara Women's University, Nara}
\affiliation{National Central University, Chung-li}
\affiliation{National United University, Miao Li}
\affiliation{Department of Physics, National Taiwan University, Taipei}
\affiliation{H. Niewodniczanski Institute of Nuclear Physics, Krakow}
\affiliation{Nippon Dental University, Niigata}
\affiliation{Niigata University, Niigata}
\affiliation{University of Nova Gorica, Nova Gorica}
\affiliation{Novosibirsk State University, Novosibirsk}
\affiliation{Osaka City University, Osaka}
\affiliation{Panjab University, Chandigarh}
\affiliation{RIKEN BNL Research Center, Upton, New York 11973}
\affiliation{Saga University, Saga}
\affiliation{University of Science and Technology of China, Hefei}
\affiliation{Seoul National University, Seoul}
\affiliation{Sungkyunkwan University, Suwon}
\affiliation{School of Physics, University of Sydney, NSW 2006}
\affiliation{Excellence Cluster Universe, Technische Universit\"at M\"unchen, Garching}
\affiliation{Toho University, Funabashi}
\affiliation{Tohoku Gakuin University, Tagajo}
\affiliation{Department of Physics, University of Tokyo, Tokyo}
\affiliation{Tokyo Metropolitan University, Tokyo}
\affiliation{Tokyo University of Agriculture and Technology, Tokyo}
\affiliation{IPNAS, Virginia Polytechnic Institute and State University, Blacksburg, Virginia 24061}
\affiliation{Yonsei University, Seoul}
  \author{K.-F.~Chen}\affiliation{Department of Physics, National Taiwan University, Taipei} 
  \author{W.-S.~Hou}\affiliation{Department of Physics, National Taiwan University, Taipei} 
  \author{I.~Adachi}\affiliation{High Energy Accelerator Research Organization (KEK), Tsukuba} 
  \author{H.~Aihara}\affiliation{Department of Physics, University of Tokyo, Tokyo} 
  \author{T.~Aushev}\affiliation{\'Ecole Polytechnique F\'ed\'erale de Lausanne (EPFL), Lausanne}\affiliation{Institute for Theoretical and Experimental Physics, Moscow} 
  \author{A.~M.~Bakich}\affiliation{School of Physics, University of Sydney, NSW 2006} 
  \author{V.~Balagura}\affiliation{Institute for Theoretical and Experimental Physics, Moscow} 
  \author{A.~Bay}\affiliation{\'Ecole Polytechnique F\'ed\'erale de Lausanne (EPFL), Lausanne} 
  \author{K.~Belous}\affiliation{Institute of High Energy Physics, Protvino} 
  \author{V.~Bhardwaj}\affiliation{Panjab University, Chandigarh} 
% ...
\collaboration{The Belle Collaboration}

\maketitle

\tighten

A host of new charmonium-like mesons have been discovered
recently that do not seem to fit into the conventional
$c\overline{c}$ spectrum. 
Possible interpretations for these
exotic states include multiquark states,
$c\overline{q}$--$\overline{c}q$ mesonic-molecules
(where $q$ represents a $u$-,
$d$- or $s$-quark), or $c\overline{c}g$ ``hybrids" (where $g$ is a
gluon). The narrow $X(3872)$~\cite{Choi:2003ue}, 
and various vector mesons, in particular,
the broad $Y(4260)$~\cite{Aubert:2005rm,Coan:2006rv,:2007sj} 
resonance, were revealed by their
dipion transitions to $J/\psi$ (and $\psi^\prime$).
By analogy with the $Y(4260)$ discovery, it was suggested that a
hidden-beauty counterpart, denoted $Y_b$~\cite{Hou:2006it}, could
be sought in radiative return events from $e^+e^-$ at center-of-mass 
energy on the $\Upsilon(10860)$ peak, or by an energy scan above
it.

Analyzing a data sample recorded by the Belle detector at a single
energy near the $\Upsilon(10860)$ resonance peak, no
unusual structure was observed below the peak. Instead,
anomalously large $\Upsilon(1S)\pi^+ \pi^-$ and $\Upsilon(2S)\pi^+
\pi^-$ production rates were observed~\cite{Abe:2007tk}. If these
signals are attributed entirely to dipion transitions from the
$\Upsilon(10860)$ resonance, the corresponding partial widths would be 
two orders of magnitude larger than those for
corresponding transitions from the $\Upsilon(2S)$, $\Upsilon(3S)$,
and $\Upsilon(4S)$ states. Possible explanations include a new
nonperturbative approach~\cite{Simonov:2007bm} to  the decay
widths of dipion transitions of heavy quarkonia, the presence of
final state interactions~\cite{Meng:2007tk}, or the existence of a
tetraquark state~\cite{Karliner:2008rc,Ali:2009pi}.
A measurement of the energy
dependence of the cross sections for $e^+e^- \to \Upsilon(nS)\pi^+
\pi^-$ ($n=1,2,3$) in and above the $\Upsilon(10860)$ energy
region provides more information for exploring 
these models.



Here we report the observation of the production of $e^+e^-
\to \Upsilon(1S)\pi^+\pi^-$, $\Upsilon(2S)\pi^+\pi^-$, and
$\Upsilon(3S)\pi^+\pi^-$ at $\sqrt{s}\simeq 10.83$, $10.87$,
$10.88$, $10.90$, $10.93$, $10.96$, and $11.02$ GeV
based on a previous data sample of 21.7 fb$^{-1}$
already reported in Ref.~\cite{Abe:2007tk} and 
a new 8.1 fb$^{-1}$ data sample at the other energies, both
collected with the Belle detector 
at the KEKB $e^+e^-$ collider~\cite{ref:KEKB}. 
In addition, nine scan points with an integrated luminosity of
$\simeq$ 30 pb$^{-1}$ each, collected in the range of
$\sqrt{s}=10.80$--$11.02$ GeV, are used to obtain the hadronic
line shape of the $\Upsilon(10860)$.

\begin{table*}[htbp]
\caption{
Center-of-mass energy ($\sqrt{s}$), integrated luminosity ($\mathcal{L}$),
signal yield ($N_s$), reconstruction efficiency, and measured cross section ($\sigma$)
for $e^+e^- \to \Upsilon(1S)\pi^+\pi^-$, $\Upsilon(2S)\pi^+\pi^-$, and $\Upsilon(3S)\pi^+\pi^-$.
Due to a negative yield, an upper limit at 90\% confidence level 
for $\sigma[e^+e^-\to\Upsilon(3S)\pi^+\pi^-]$ at $\sqrt{s}=10.9555$ GeV is given.
} \label{tab:xsec_summary}
\begin{center}
\footnotesize
\begin{tabular}{cr|rcc|rcc|rcc}
\hline
\hline
    & &\multicolumn{3}{c|}{$e^+e^-\to\Upsilon(1S)\pi^+\pi^-$} & \multicolumn{3}{c|}{$e^+e^-\to\Upsilon(2S)\pi^+\pi^-$} & \multicolumn{3}{c}{$e^+e^-\to\Upsilon(3S)\pi^+\pi^-$}\\
$\sqrt{s}$(GeV) & $\mathcal{L}$(fb$^{-1}$) & $N_s$~~~ & Eff.(\%) & $\sigma$(pb)~~~~~~ & $N_s$~~~ & Eff.(\%) & $\sigma$(pb)~~~~~~ & $N_s$~~~ & Eff.(\%) & $\sigma$(pb)~~~~~~ \\
\hline
10.8255 &  1.73~~ & $10.6_{-3.3}^{+4.0}$ & 43.8 & $0.56_{-0.18}^{+0.21}\pm0.06$   & $24.0_{-4.9}^{+5.6}$ & 34.9 & $2.05_{-0.42}^{+0.48}\pm0.24$      & $ 1.8_{-1.1}^{+1.8}$ & 20.5 & $ 0.23_{-0.14}^{+0.23}\pm0.03$ \\
10.8805 &  1.89~~ & $43.4_{-6.5}^{+7.2}$ & 43.1 & $2.14_{-0.32}^{+0.36}\pm0.15$   & $68.8_{-8.3}^{+9.0}$ & 35.4 & $5.31_{-0.64}^{+0.69}\pm0.59$      & $14.9_{-3.7}^{+4.3}$ & 24.5 & $ 1.47_{-0.37}^{+0.43}\pm0.18$ \\
10.8955 &  1.46~~ & $26.2_{-5.1}^{+5.8}$ & 43.2 & $1.68_{-0.33}^{+0.37}\pm0.13$   & $45.4_{-6.7}^{+7.4}$ & 35.6 & $4.53_{-0.67}^{+0.74}\pm0.51$      & $10.3_{-3.1}^{+3.7}$ & 25.7 & $ 1.26_{-0.38}^{+0.45}\pm0.15$ \\
10.9255 &  1.18~~ & $11.1_{-3.3}^{+4.0}$ & 42.6 & $0.89_{-0.27}^{+0.32}\pm0.08$   & $ 9.7_{-3.1}^{+3.8}$ & 35.9 & $1.19_{-0.38}^{+0.47}\pm0.16$      & $ 2.9_{-1.5}^{+2.2}$ & 27.5 & $ 0.41_{-0.21}^{+0.31}\pm0.05$ \\
10.9555 &  0.99~~ & $ 3.9_{-1.9}^{+2.6}$ & 42.5 & $0.37_{-0.18}^{+0.25}\pm0.04$   & $ 2.0_{-1.3}^{+2.0}$ & 36.4 & $0.29_{-0.19}^{+0.29}\pm0.05$      & $-1.8_{-3.0}^{+2.5}$ & 29.4 & $-0.28_{-0.47}^{+0.39}\pm0.03<0.20$ \\
11.0155 &  0.88~~ & $ 4.9_{-2.1}^{+2.8}$ & 42.0 & $0.53_{-0.23}^{+0.31}\pm0.05$   & $ 5.5_{-2.4}^{+3.1}$ & 36.0 & $0.90_{-0.39}^{+0.51}\pm0.17$      & $ 4.3_{-1.9}^{+2.6}$ & 32.7 & $ 0.69_{-0.30}^{+0.42}\pm0.08$ \\
\hline
10.8670 & 21.74~~ &
$325^{+20}_{-19}$    & 37.4 & $1.61\pm0.10\pm0.12$ &
$186\pm15$           & 18.9 & $2.35\pm0.19\pm0.32$ &
$10.5^{+4.0}_{-3.3}$ &  1.5 & $1.44^{+0.55}_{-0.45}\pm0.19$ \\
\hline
\hline
\end{tabular}
\end{center}
\end{table*}

\begin{figure*}[t!]
\begin{center}
\includegraphics[width=4.50cm,height=2.0cm]{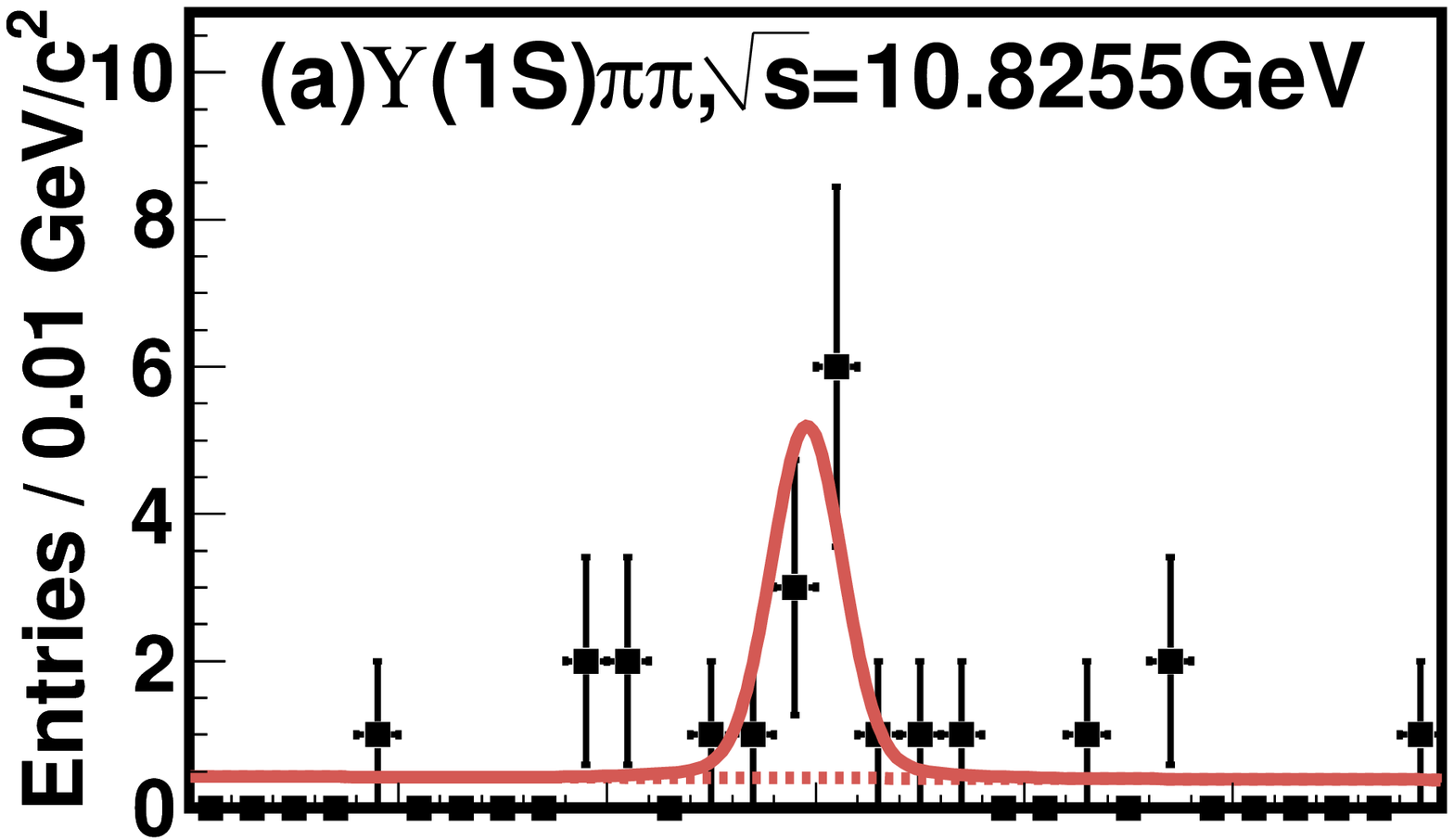}
\includegraphics[width=4.50cm,height=2.0cm]{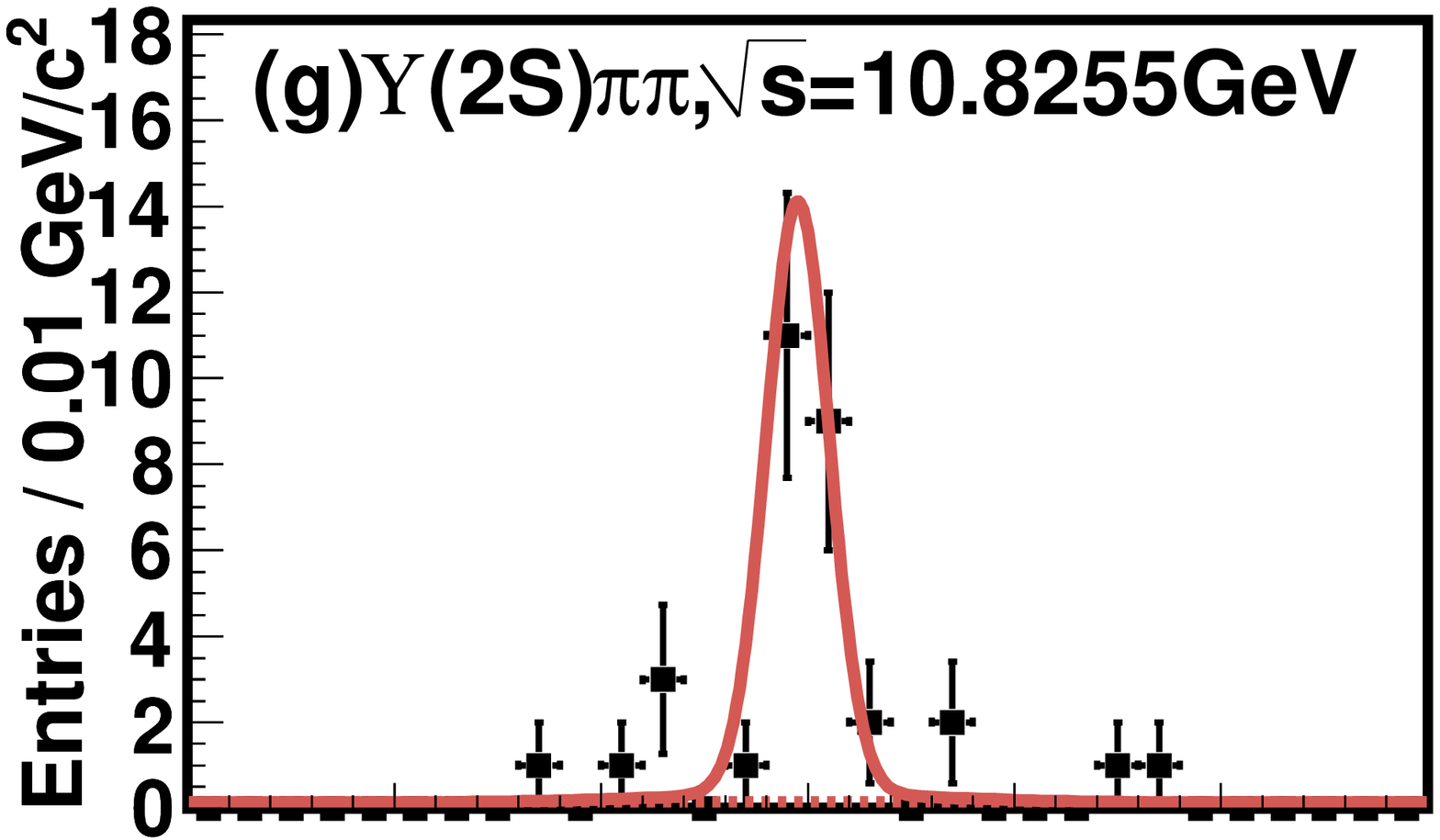}
\includegraphics[width=4.50cm,height=2.0cm]{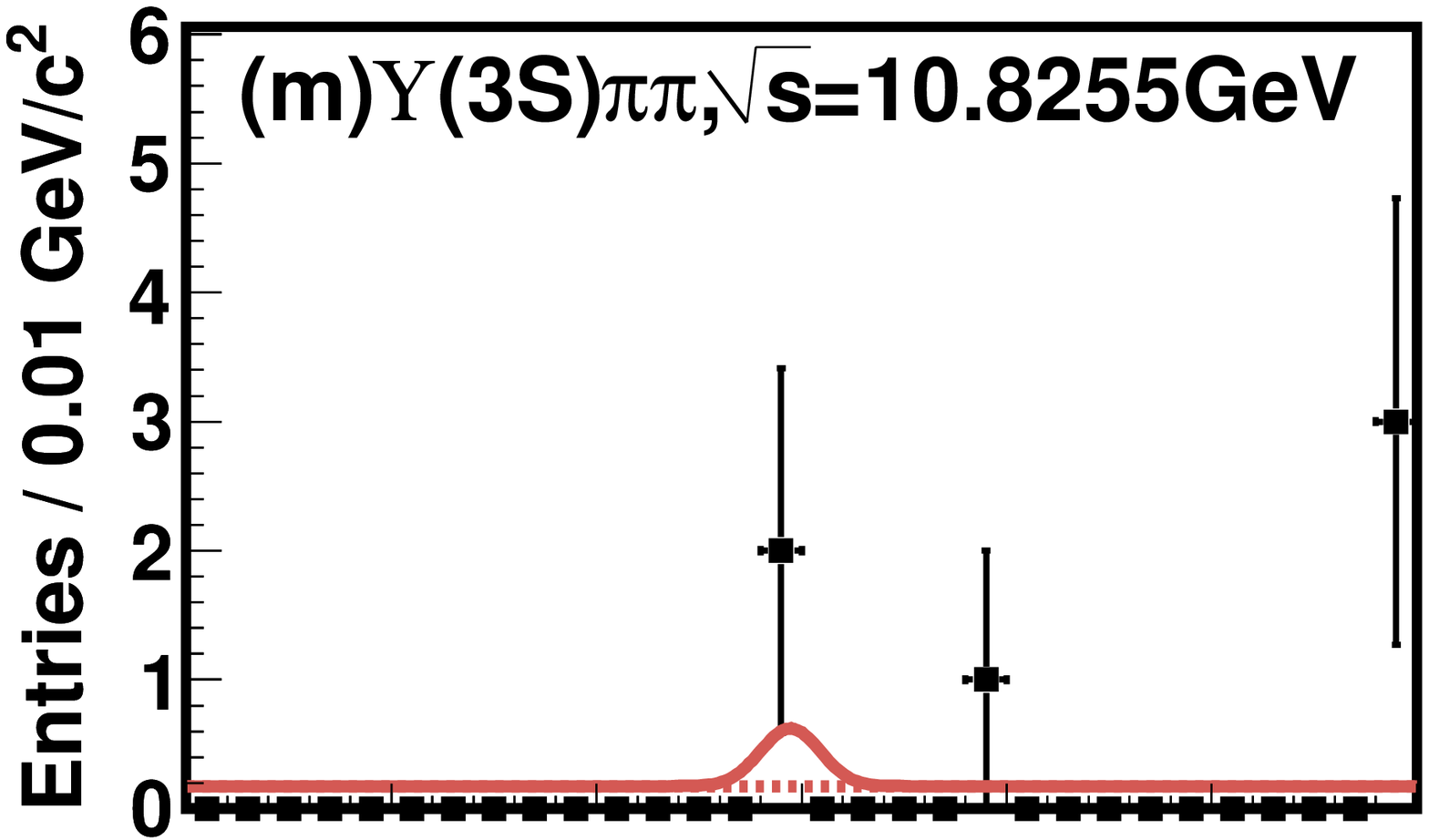}\\
\includegraphics[width=4.50cm,height=2.0cm]{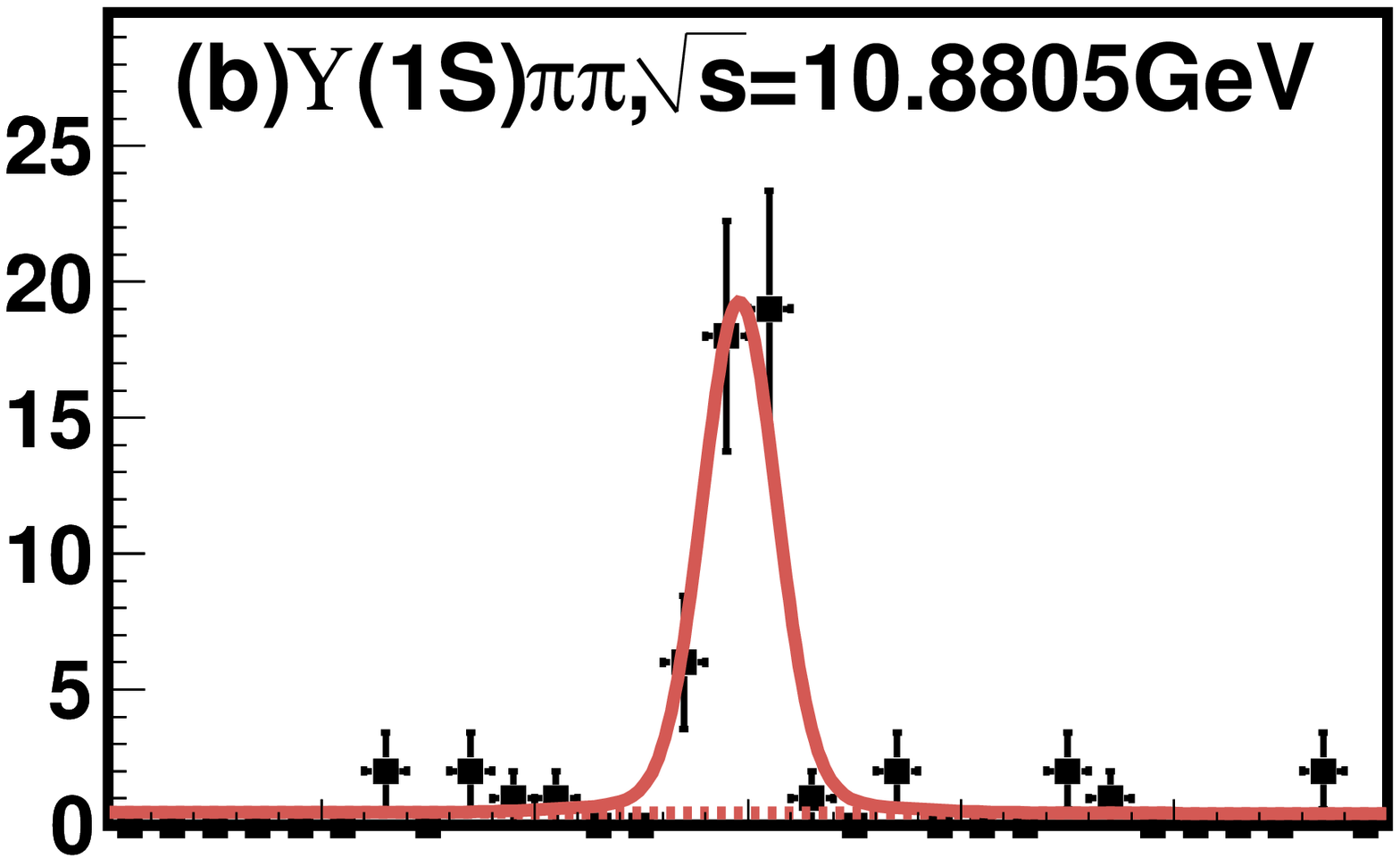}
\includegraphics[width=4.50cm,height=2.0cm]{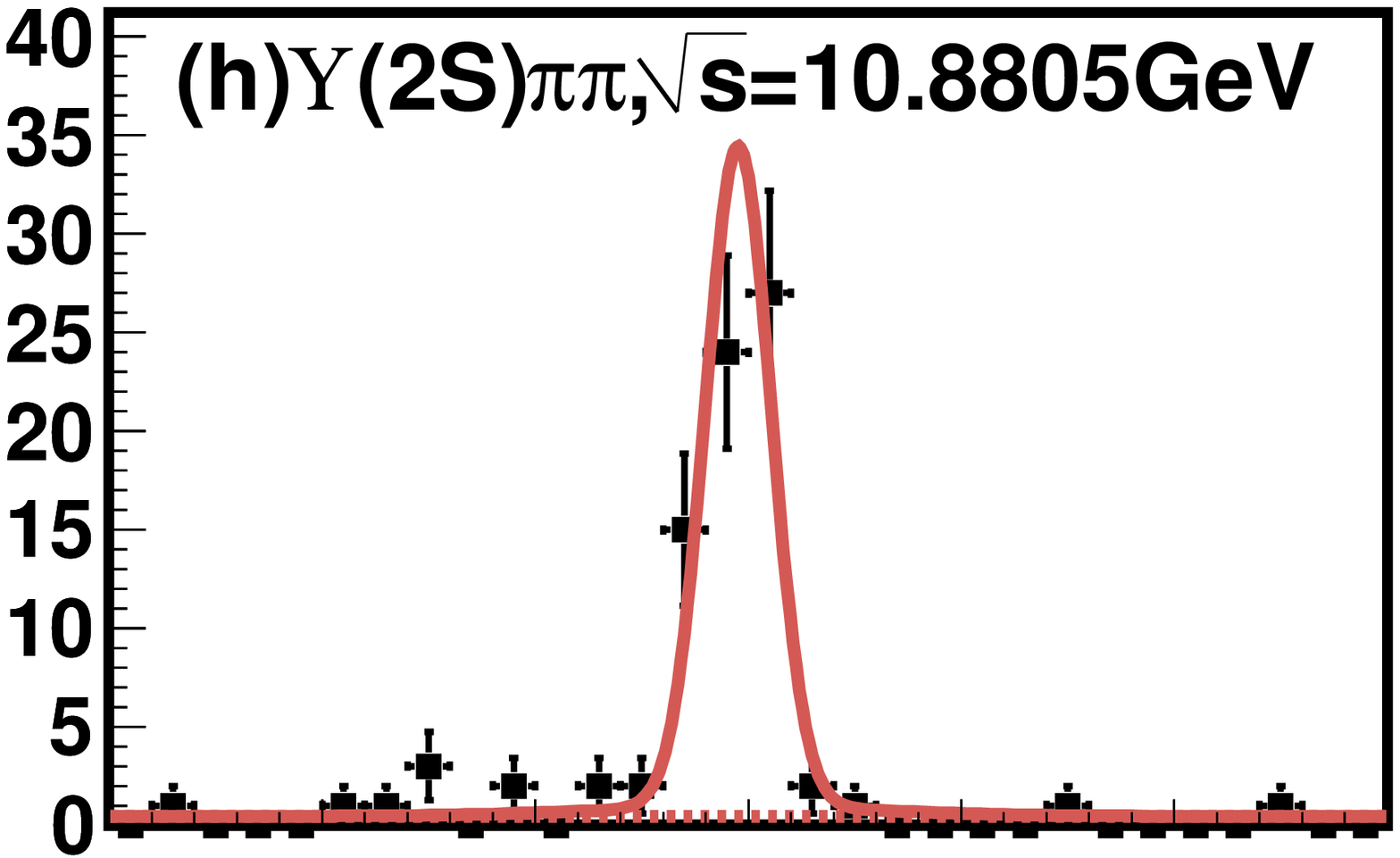}
\includegraphics[width=4.50cm,height=2.0cm]{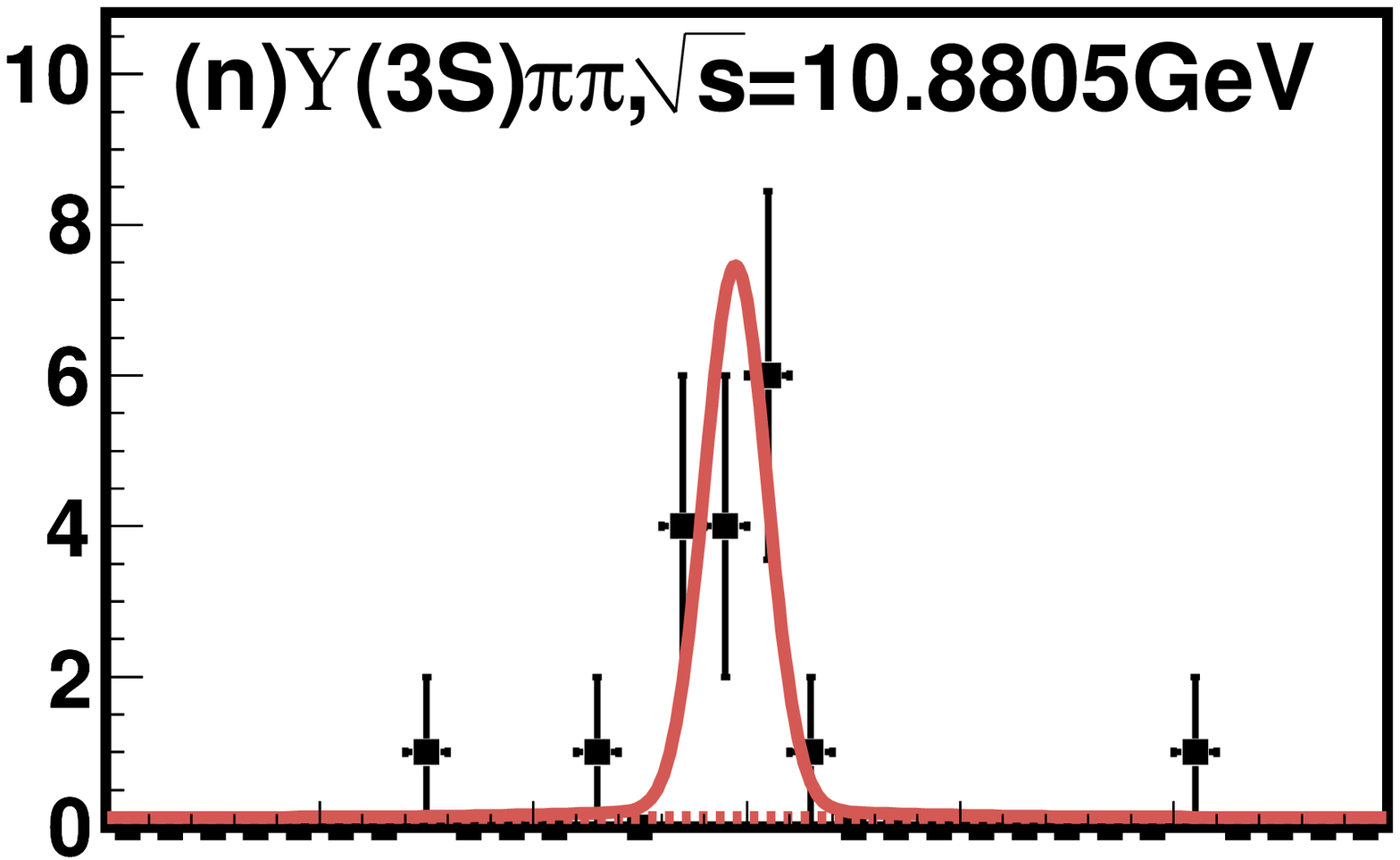}\\
\includegraphics[width=4.50cm,height=2.0cm]{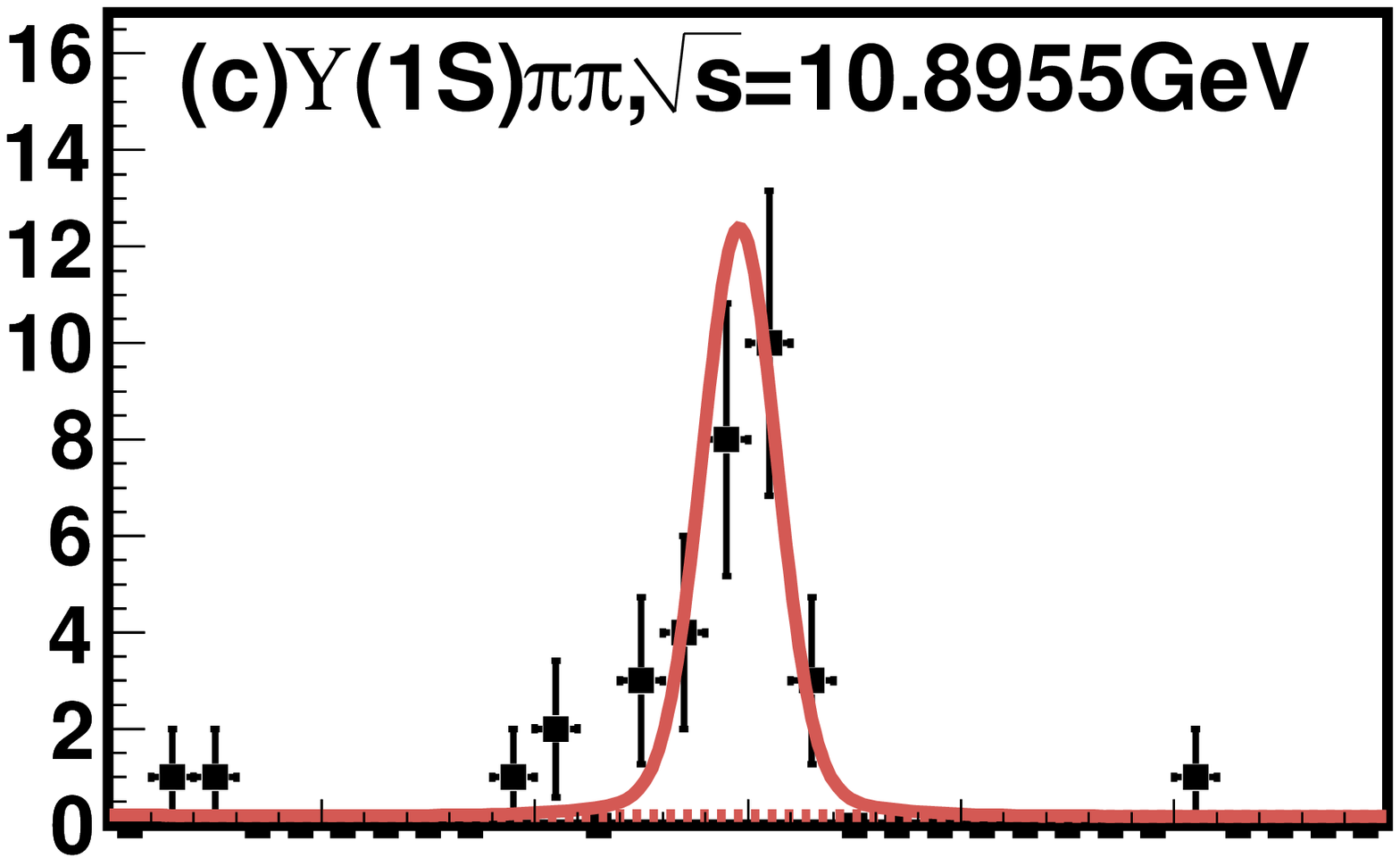}
\includegraphics[width=4.50cm,height=2.0cm]{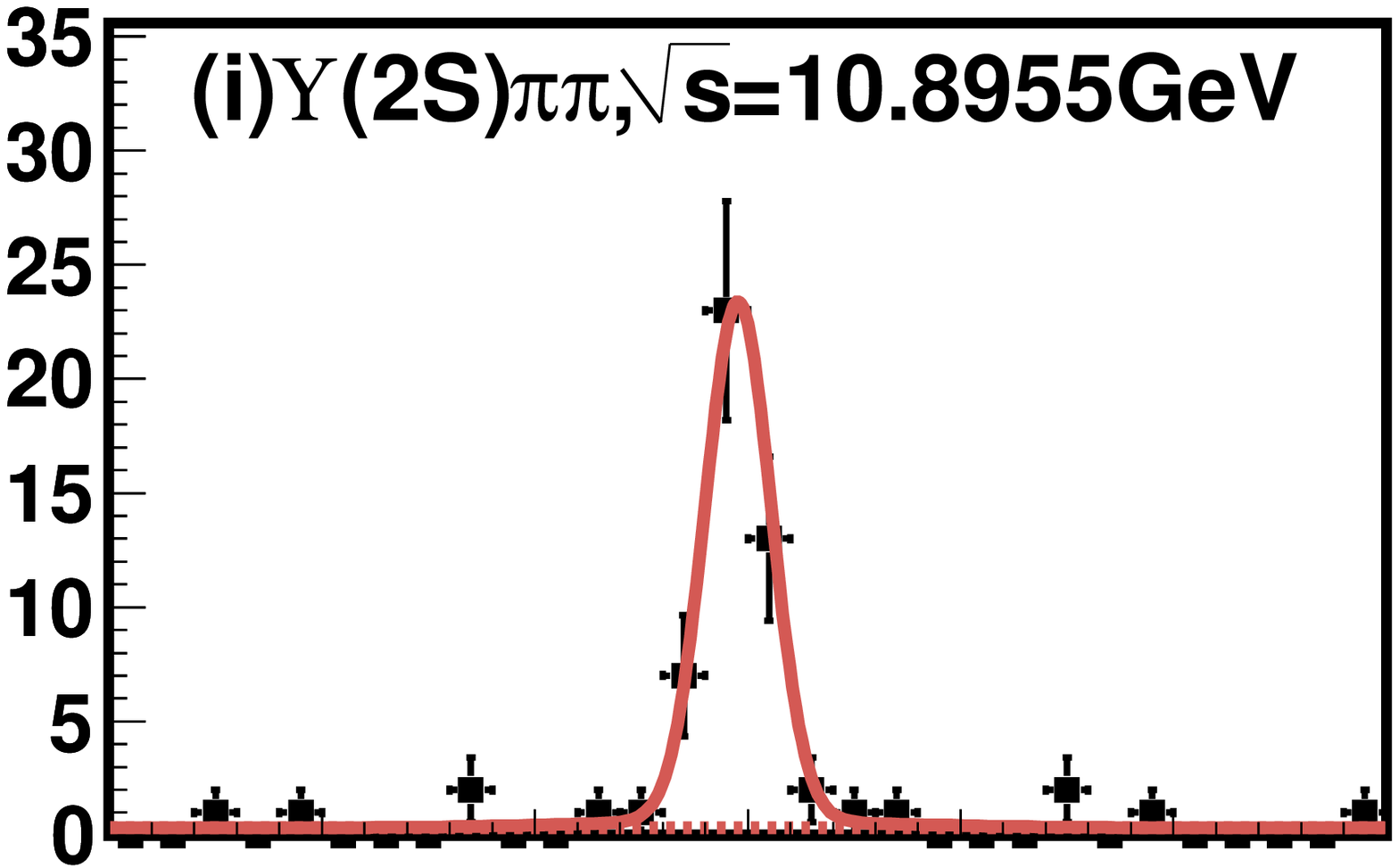}
\includegraphics[width=4.50cm,height=2.0cm]{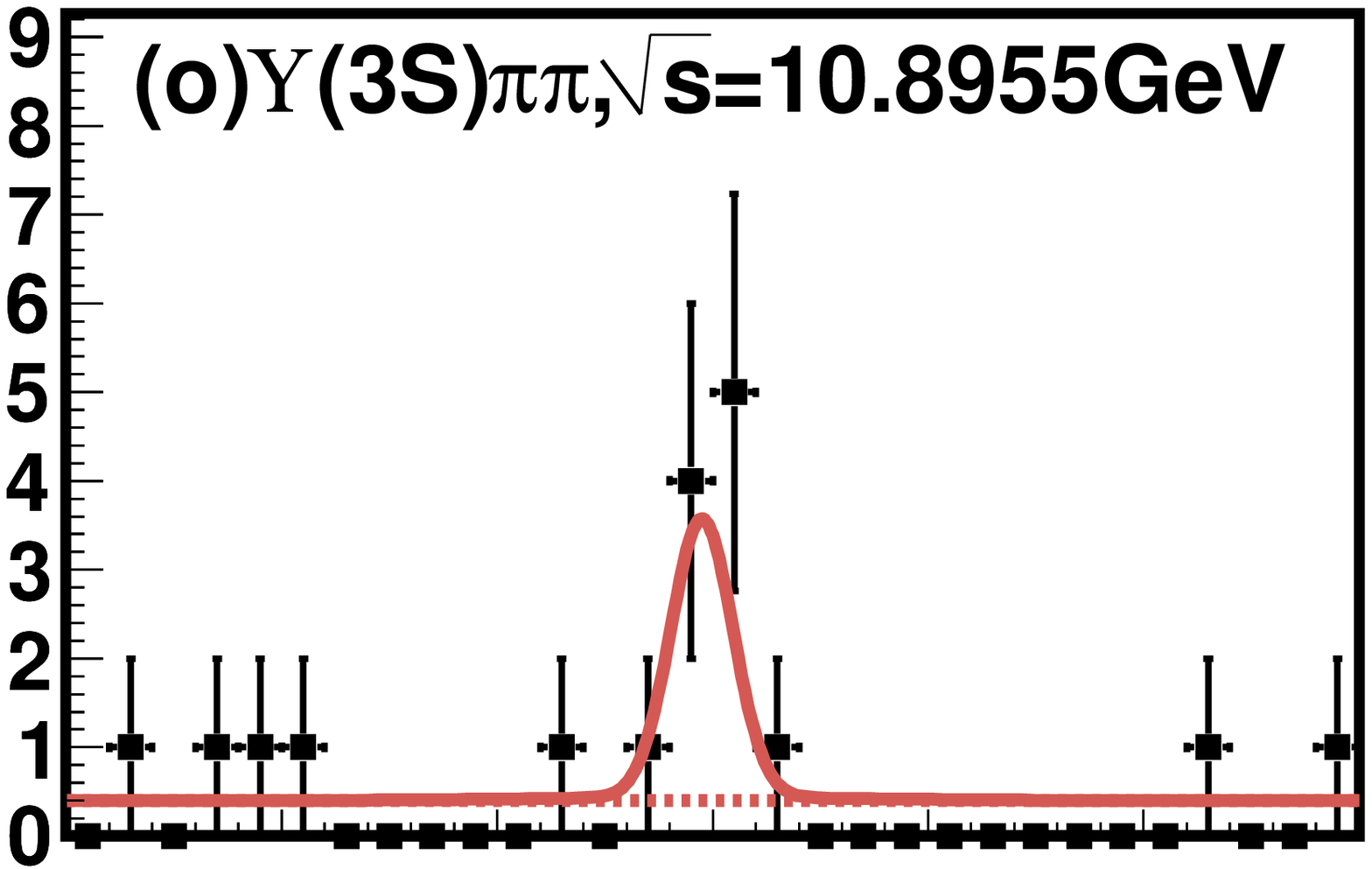}\\
\includegraphics[width=4.50cm,height=2.0cm]{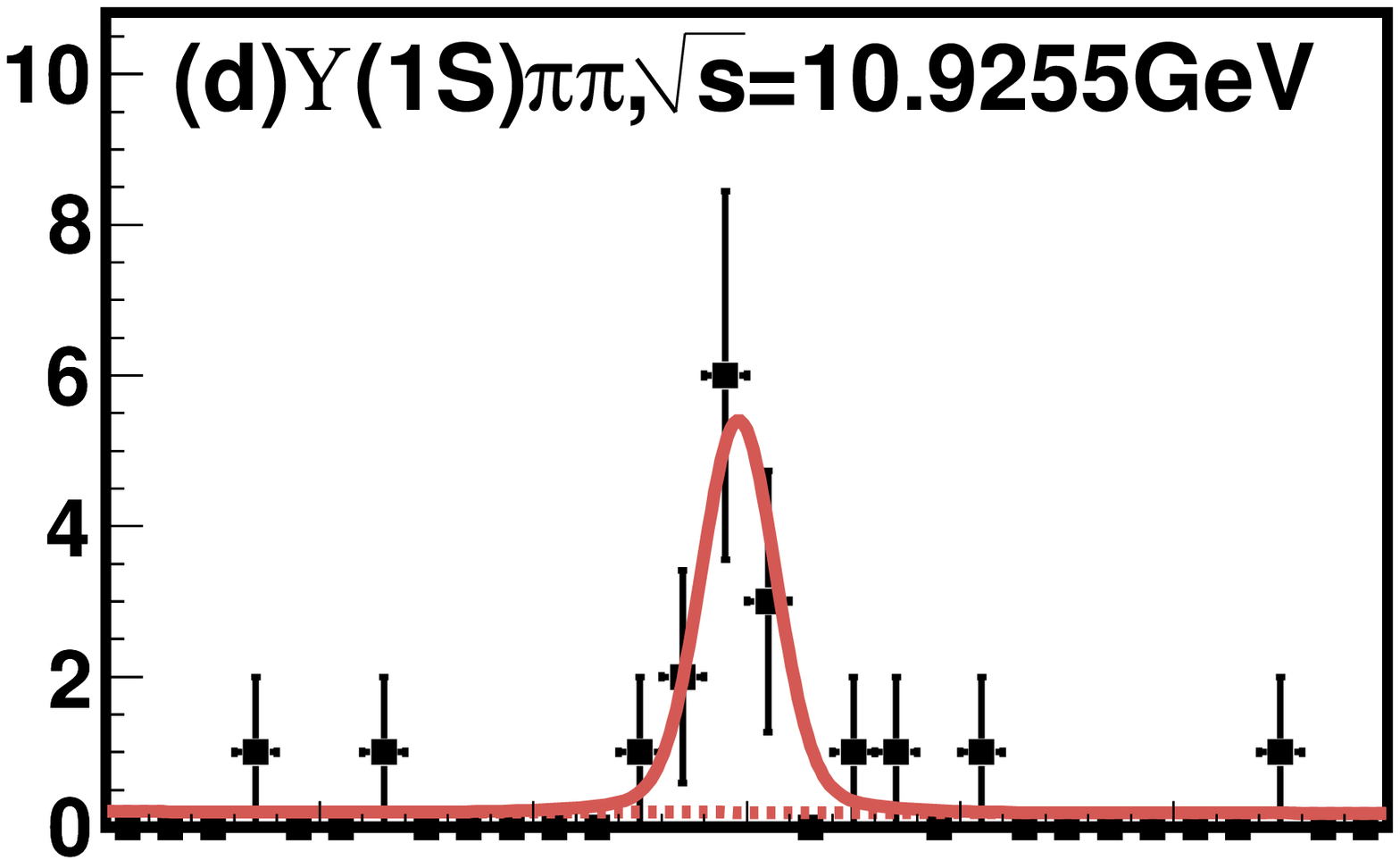}
\includegraphics[width=4.50cm,height=2.0cm]{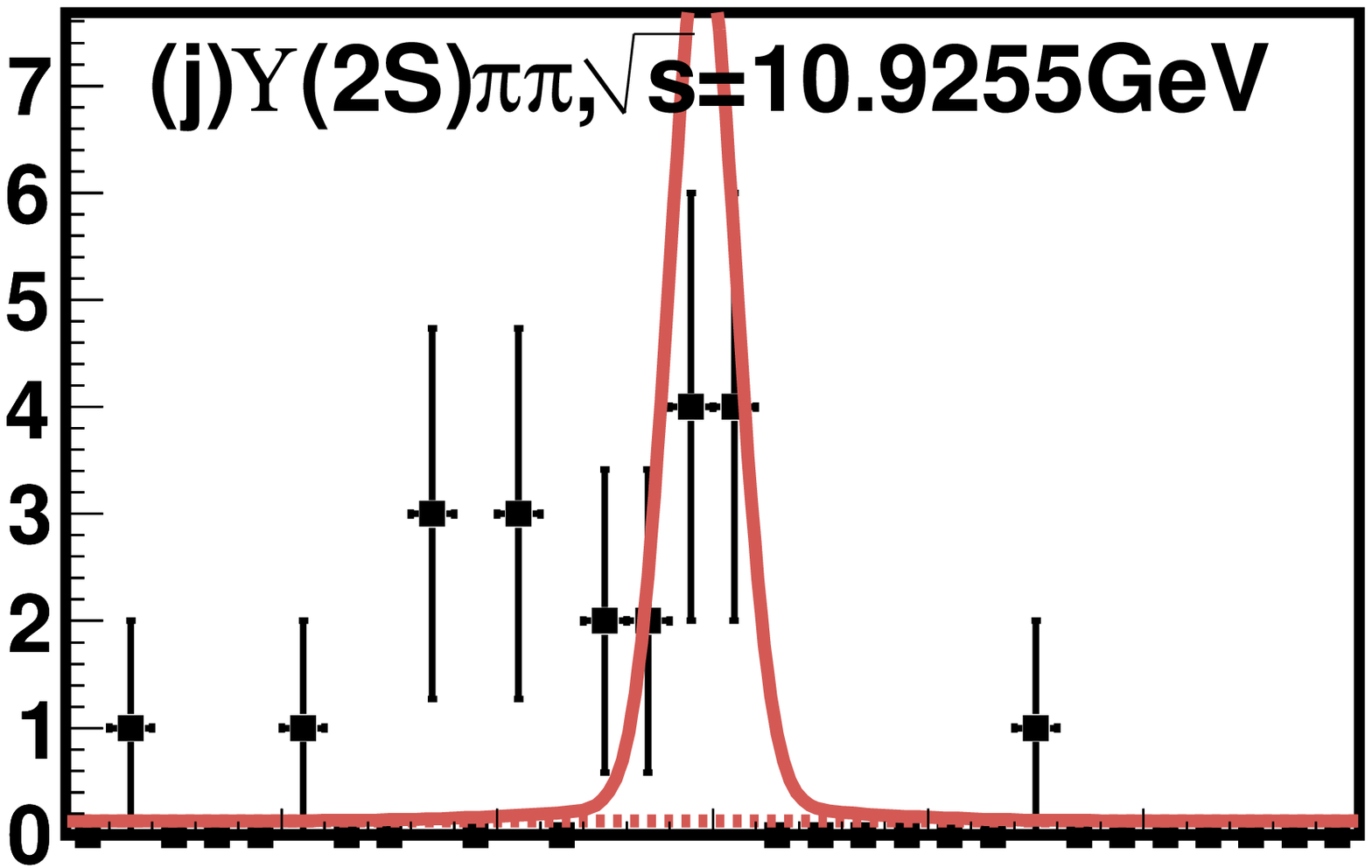}
\includegraphics[width=4.50cm,height=2.0cm]{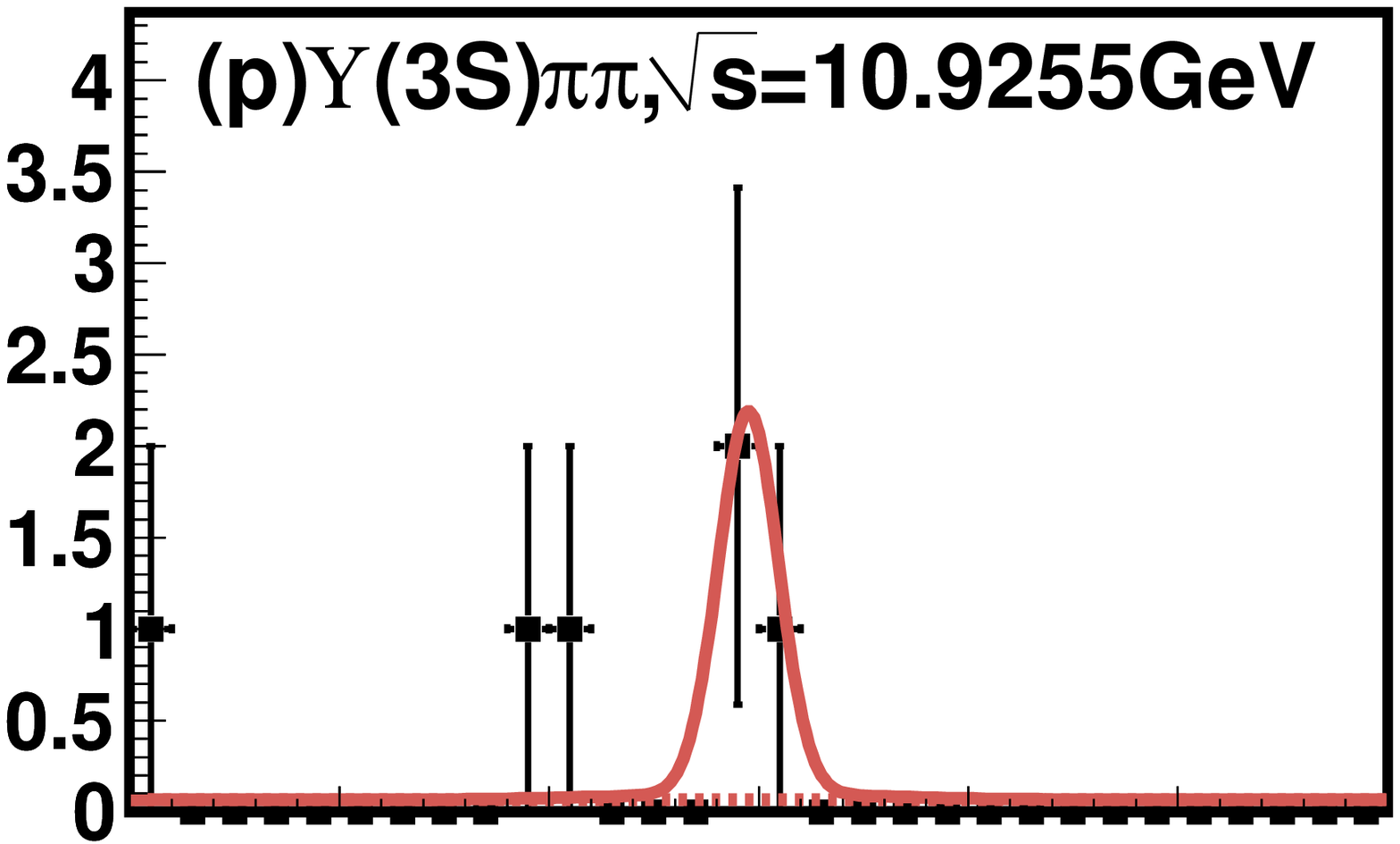}\\
\includegraphics[width=4.50cm,height=2.0cm]{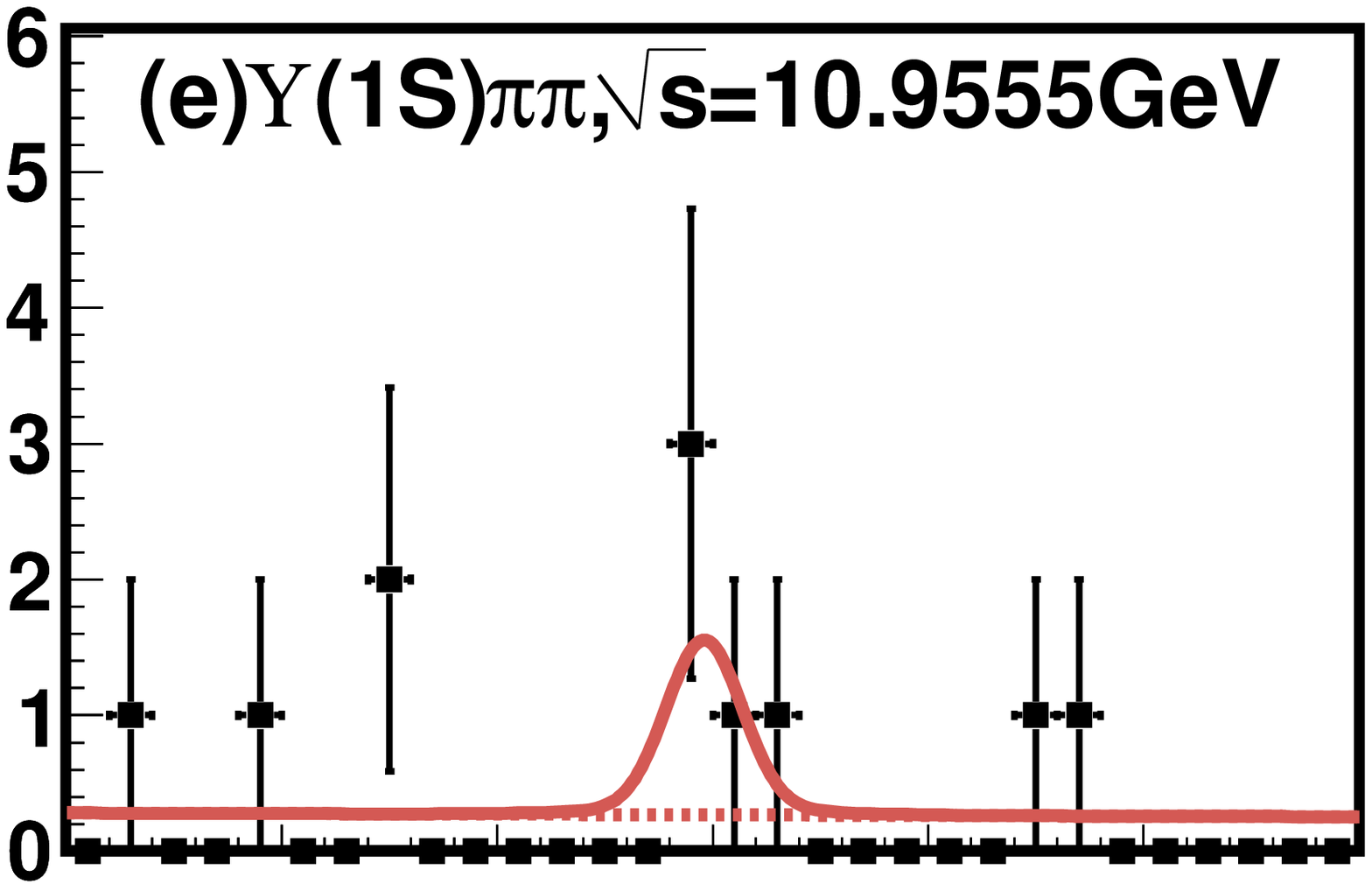}
\includegraphics[width=4.50cm,height=2.0cm]{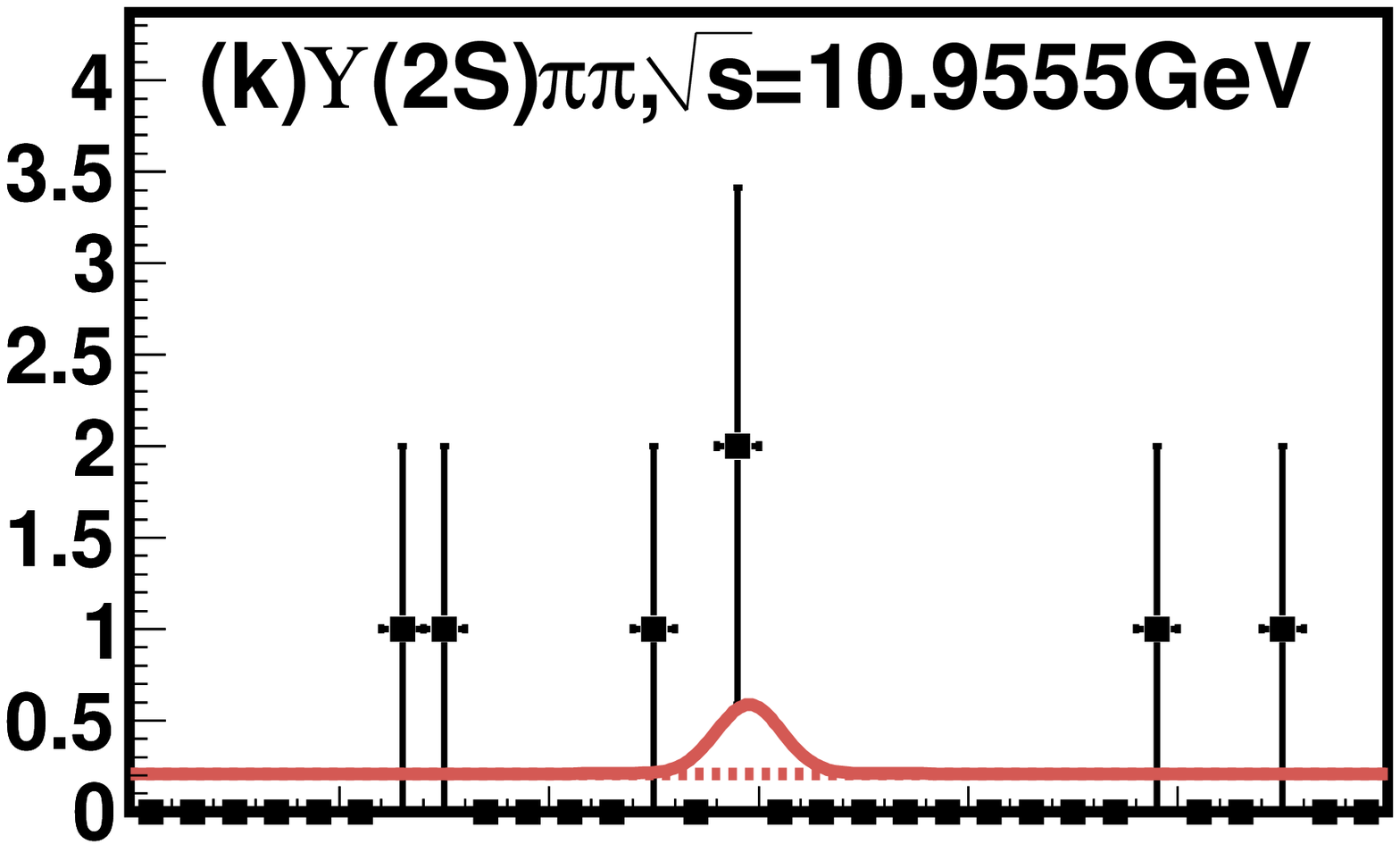}
\includegraphics[width=4.50cm,height=2.0cm]{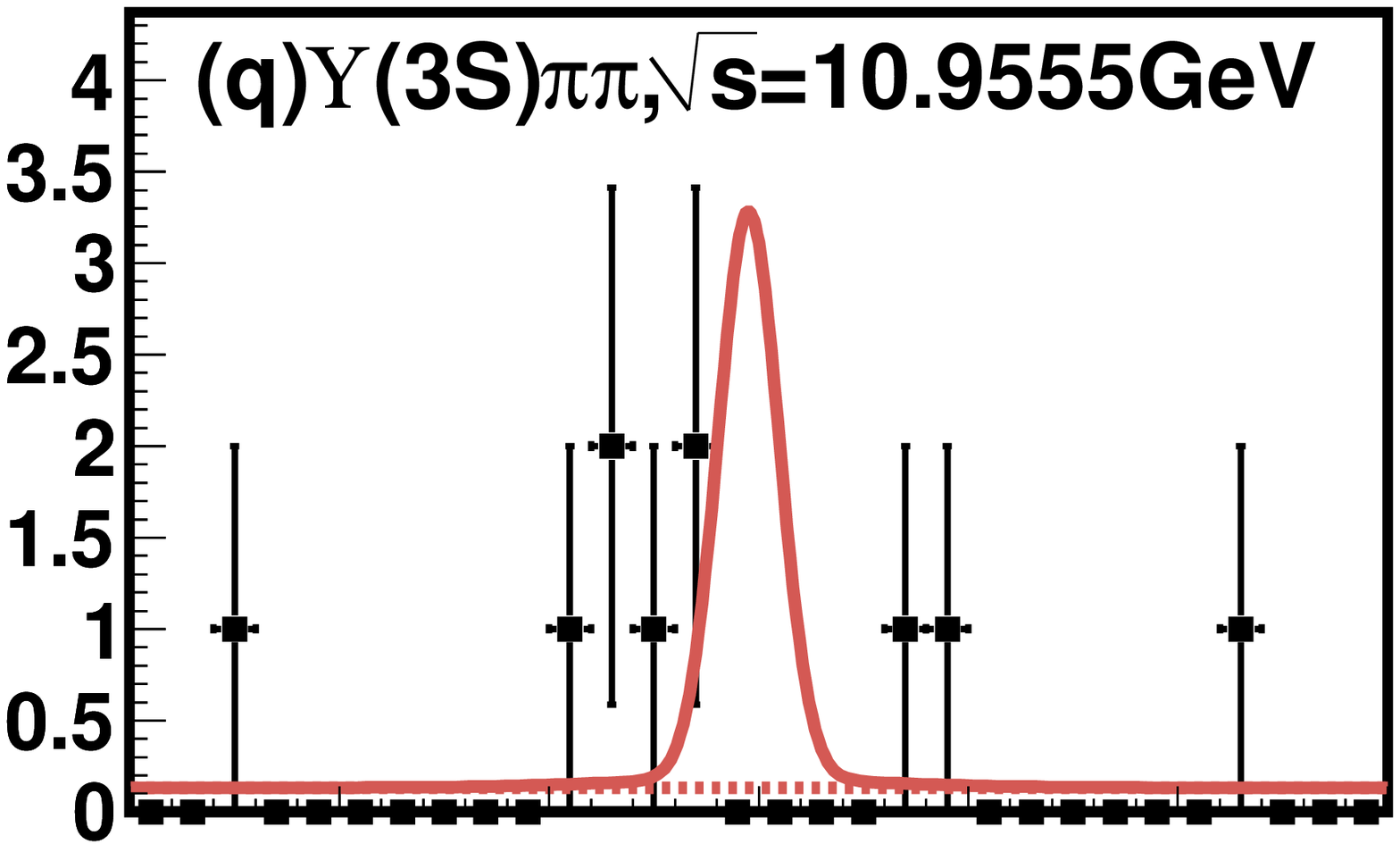}\\
\includegraphics[width=4.50cm,height=2.477cm]{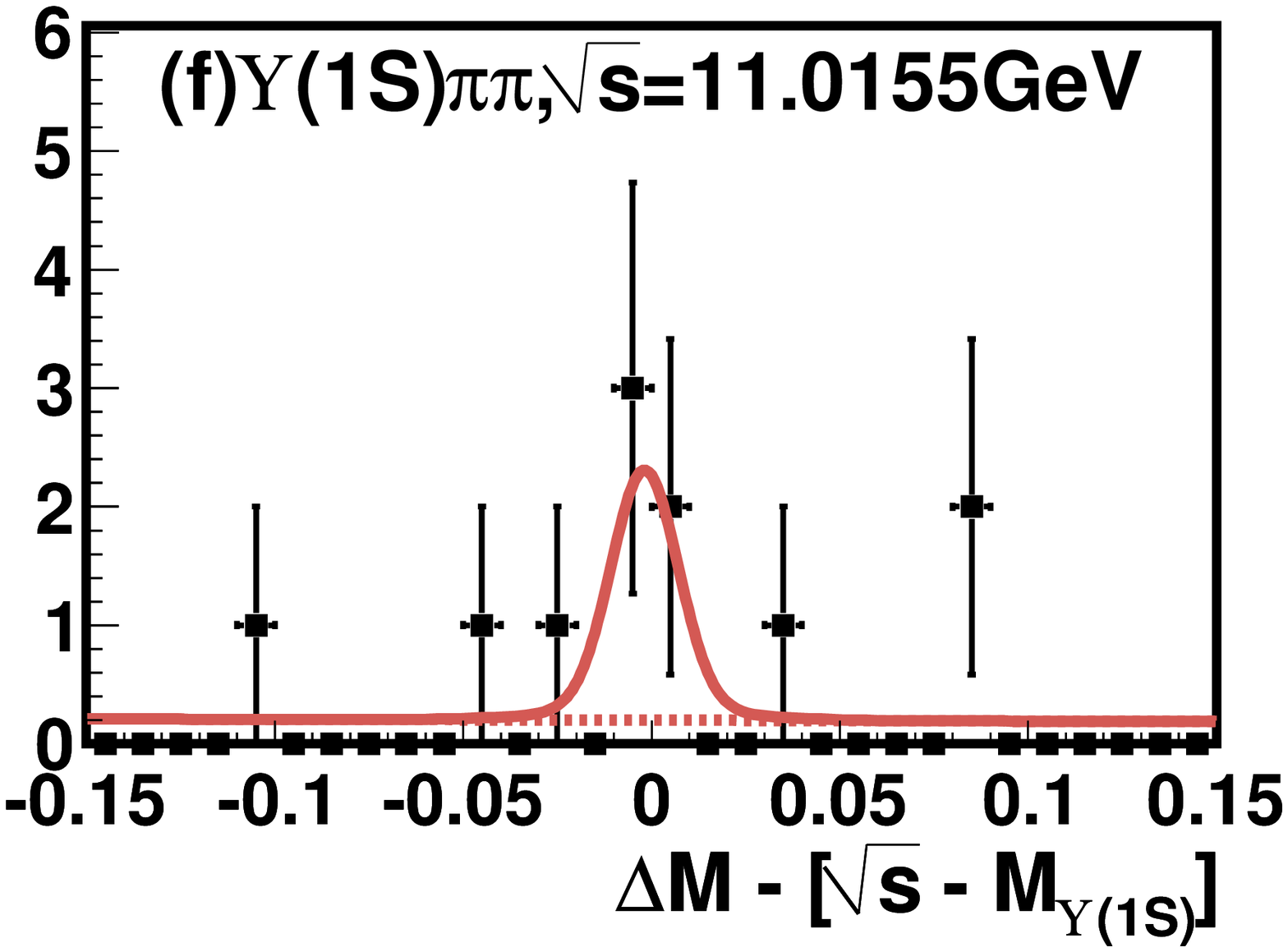}
\includegraphics[width=4.50cm,height=2.477cm]{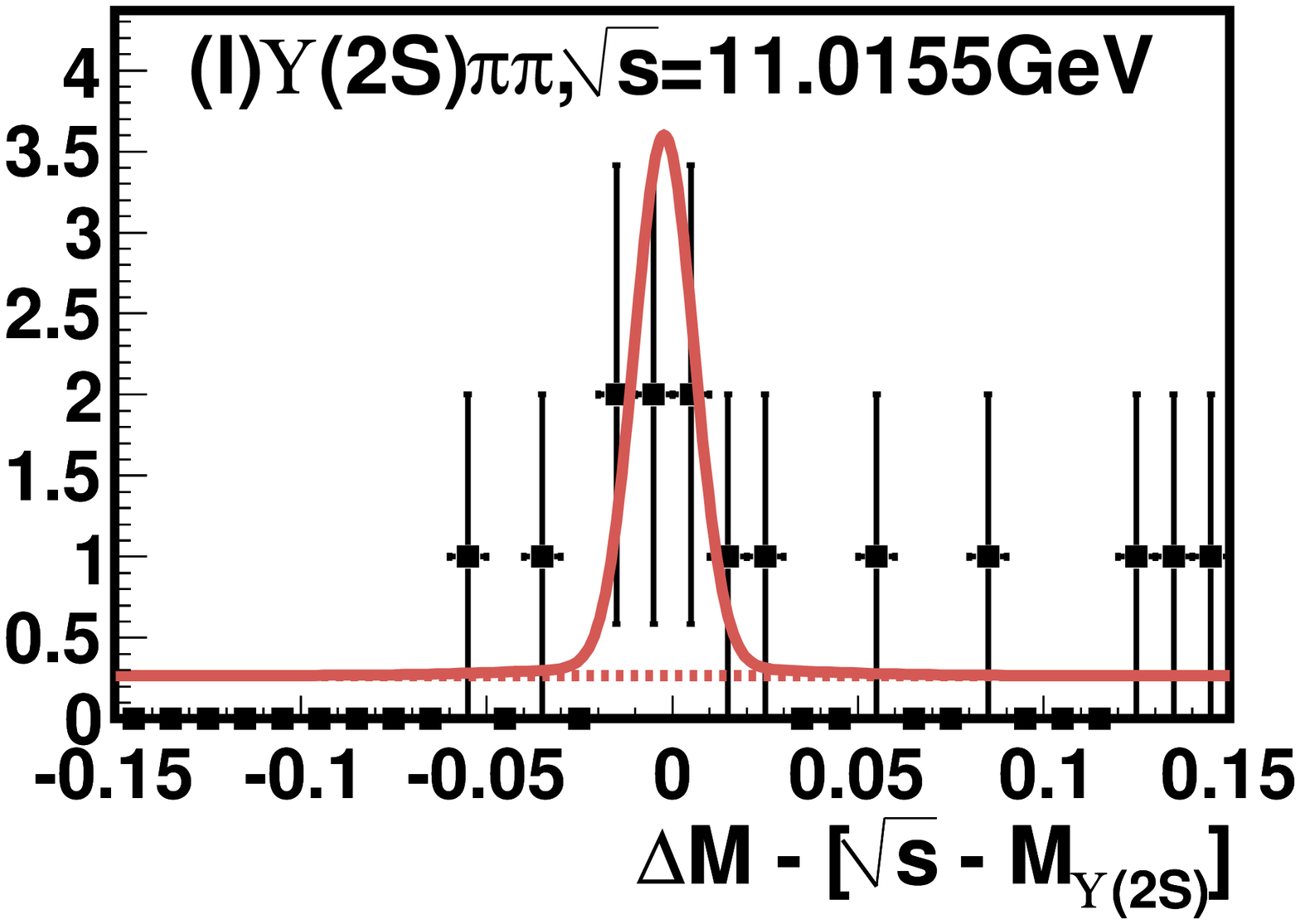}
\includegraphics[width=4.50cm,height=2.477cm]{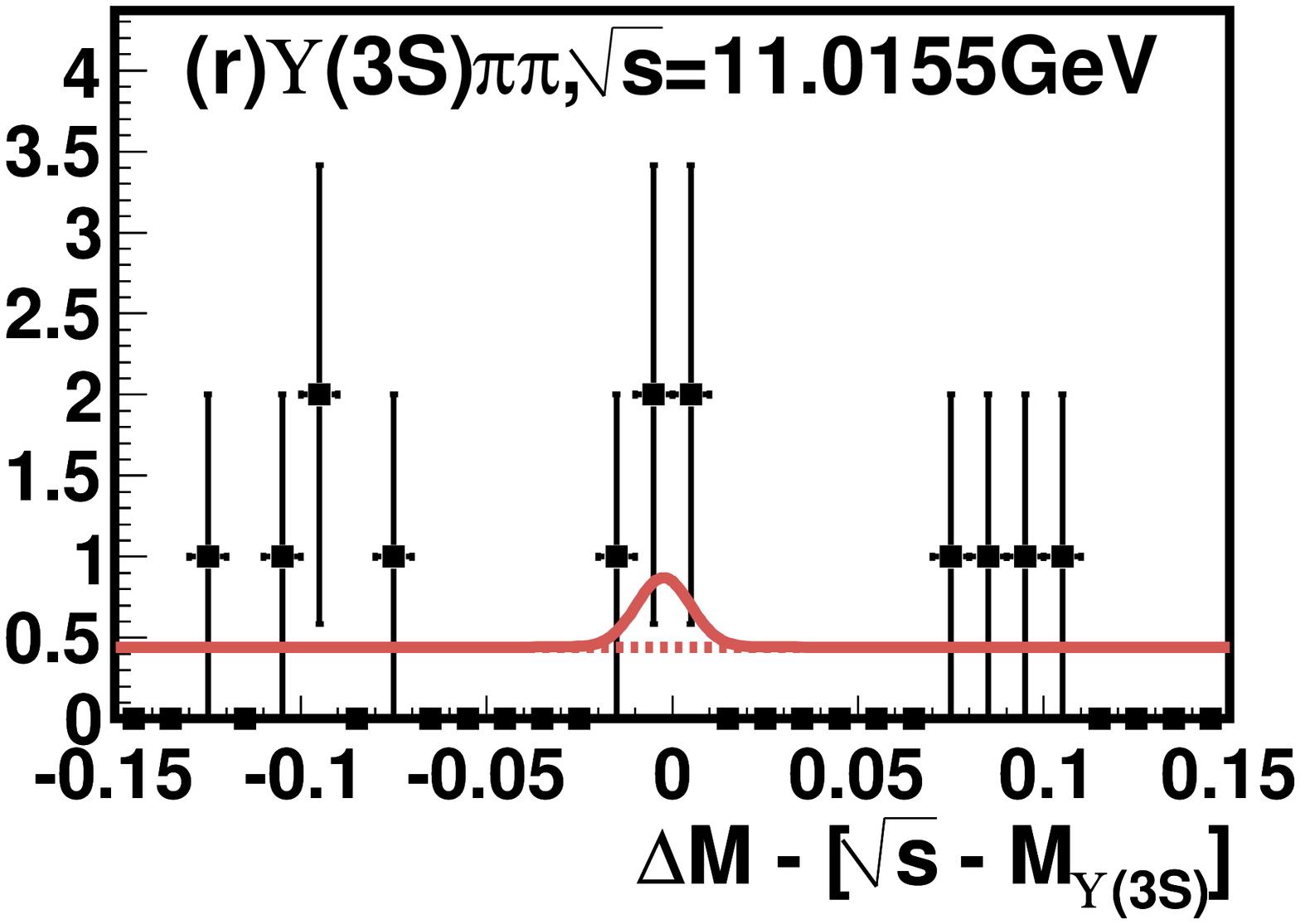}
\end{center}
\caption{The distributions of $\Delta M - [\sqrt{s} - M_{\Upsilon(nS)}]$ ($n=1,2,3$) for
 (a--f) $\Upsilon(1S)\pi^+\pi^-$,
 (g--l) $\Upsilon(2S)\pi^+\pi^-$, and
 (m--r) $\Upsilon(3S)\pi^+\pi^-$ events with the fit results superimposed.
 The six columns of plots represent the data samples collected at different energies.
 The dashed curves show the background components in the fits.}
 \label{fig:dmfit}
\end{figure*}

The Belle detector is a large-solid-angle magnetic spectrometer
that consists of a silicon vertex detector, a central drift chamber (CDC), an array of
aerogel threshold Cherenkov counters (ACC), a barrel-like arrangement of
time-of-flight scintillation counters, and an electromagnetic
calorimeter (ECL) comprised of CsI(Tl) crystals located inside a
superconducting solenoid that provides a 1.5~T magnetic field. An iron
flux-return located outside the coil is instrumented to detect $K_L^0$
mesons and to identify muons (KLM). The detector is described in detail
elsewhere~\cite{ref:belle_detector}.

Events with four well-reconstructed charged tracks and zero net charge
are selected to study $\Upsilon(nS)\pi^+\pi^-$ production. A final state
that is consistent with a $\Upsilon(nS)$ candidate and two charged pions
is reconstructed.
A $\Upsilon(nS)$ candidate is formed from two muons with opposite charge, where
the muon candidates are required
to have associated hits in the KLM detector that agree with the extrapolated
trajectory of a charged track provided by the drift chamber.
The $\Upsilon(nS)\to e^+e^-$ channels have little sensitivity due to 
the lower dielectron efficiency and 
the 20 times larger background from radiative Bhabha events.
The other two charged tracks must have a low likelihood of being electrons (the likelihood is
calculated based on the ratio of ECL shower energy to the track
momentum, $dE/dx$ from the CDC, and the ACC response);
they are then treated as pion candidates.
This requirement suppresses the background from
$e^+e^- \to \mu^+\mu^-\gamma \to
\mu^+\mu^-e^+e^-$ with a photon conversion.
The cosine of the
opening angle between the $\pi^+$ and $\pi^-$
momenta in the laboratory frame is required to be less than 0.95.
The four-track invariant mass must satisfy $|M(\mu^+\mu^-\pi^+\pi^-)-\sqrt{s}|<150$ MeV/$c^2$.
The trigger efficiency for four-track events satisfying these criteria is very close to 100\%.

The kinematic variable $\Delta M$, defined by the difference between
$M(\mu^+\mu^-\pi^+\pi^-)$ and $M(\mu^+\mu^-)$,
is used to identify the signal candidates.
Sharp signal peaks will occur at $\Delta M = \sqrt{s} - M_{\Upsilon(nS)}$.
The candidate events are separated into three distinct regions
defined by $|\Delta M - [\sqrt{s} - M_{\Upsilon(nS)}]| < 150$ MeV/$c^2$ for $n = 1$, $2$, and $3$,
and signal yields are extracted from an unbinned extended maximum likelihood fit to
the $\Delta M$ distribution within each region.
The likelihood function for each fit is defined as
\begin{equation*}
L(N_s,N_b) = {e^{-(N_s+N_b)} \over {N!}}
 \prod_{i=1}^N [ N_s \cdot P_s (\Delta M_i) + N_b \cdot P_b (\Delta M_i) ]~,
\end{equation*}
where $N_s$ ($N_b$) denotes the yield for signal (background), and
$P_s$ ($P_b$) is the signal (background) probability density
function (PDF). The signal is modelled by a sum of two Gaussians
while the background is approximated by a linear function. The
Gaussians parameterized for the signal PDF are fixed from the
Monte Carlo (MC) simulation at each energy point. We fit 18
$\Delta M$ distributions simultaneously with common corrections
to the mean and width of the signal Gaussians.
The widths are found to be around $19\pm8$\% higher than the
expectations given by the MC simulations. 
Figure~\ref{fig:dmfit} shows these $\Delta M$ distributions with fit results
superimposed; the other three $\Delta M$ distributions for the data collected 
at $\sqrt{s} \approx 10.867$ GeV are included in Ref.~\cite{Abe:2007tk}.


The measured signal yields, reconstruction efficiencies,
integrated luminosity, and the production cross sections,
as well as the results from the previous publication~\cite{Abe:2007tk} for
the data sample collected at $\sqrt{s}=10.867$ GeV,
are summarized in Table~\ref{tab:xsec_summary}.
The efficiencies for $\Upsilon(3S)\pi^+\pi^-$ are much improved compared to Ref.~\cite{Abe:2007tk}
as the inefficient selection criterion $\theta_{\rm max}<175^\circ$ has been removed,
where $\theta_{\rm max}$ is the
maximum opening angle between any pair of charged tracks in the center-of-mass frame.
This channel is clearly seen for the first time.
The selection criteria already constrain the 
reconstructed total energy to be very close to the input beam energy.
The contribution from radiative return events is expected to be very low.
The observed peaks in $\Delta M$ distributions 
agree with MC expectations without initial state radiation.


For the cross section measurements, systematic uncertainties are
dominated by the $\Upsilon(nS)\to\mu^+\mu^-$ branching fractions,
reconstruction efficiencies, and PDF parameterization for the fits.
Uncertainties of 2.0\%, 8.8\%, and 9.6\% for the $\Upsilon(1S)$,
$\Upsilon(2S)$, and $\Upsilon(3S)\to\mu^+\mu^-$ branching
fractions are included, respectively. For the
$\Upsilon(1S)\pi^+\pi^-$ and $\Upsilon(2S)\pi^+\pi^-$ modes, the
reconstruction efficiencies are obtained from MC simulations using
the observed $M(\pi^+\pi^-)$ and $\cos\theta_{\rm Hel}$
(the angle between the $\pi^-$ and $\Upsilon(10860)$ momenta in the $\pi^+\pi^-$ rest frame)
distributions in our previous publication as inputs~\cite{Abe:2007tk}. 
The uncertainties associated with these distributions give rise to
2.7\%--4.5\% and 1.9\%--4.2\% uncertainties for the
$\Upsilon(1S)\pi^+\pi^-$ and $\Upsilon(2S)\pi^+\pi^-$
efficiencies, respectively. The ranges on the uncertainty arise
from the energy dependence of the $\pi^+\pi^-$ system. We use the model of
Ref.~\cite{Brown:1975dz} as well as a phase space model as inputs
for $\Upsilon(3S)\pi^+\pi^-$ measurements; the differences in
acceptance are included as systematic uncertainties. The
uncertainties from the PDF parameterization are estimated either by
replacing the signal PDF with a sum of three Gaussians, or by
replacing the background PDF with a second-order polynomial. The
differences between these alternative fits and the nominal results
are taken as the systematic uncertainties. Other uncertainties
include: tracking efficiency ($1\%$ per charged track), muon
identification (0.5\% per muon candidate), electron rejection for
the charged pions (0.1--0.2\% per pion), trigger efficiencies
(0.1--5.2\%), and integrated luminosity (1.4\%). The uncertainties
from all sources are added in quadrature. The total systematic
uncertainties are 7\%--11\%, 11\%--16\%, and 12\%--14\% for the
$\Upsilon(1S)\pi^+\pi^-$, $\Upsilon(2S)\pi^+\pi^-$, and
$\Upsilon(3S)\pi^+\pi^-$ channels, respectively.

\begin{figure}[t!]
\begin{center}
\includegraphics[width=8.5cm,height=3.5cm]{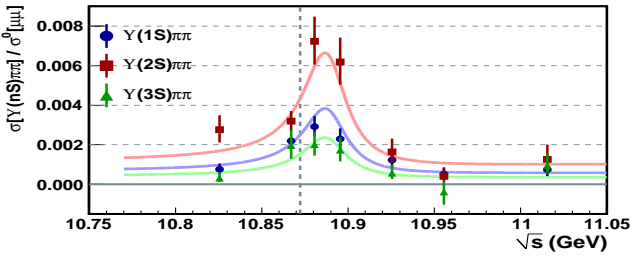}
\end{center}
\caption{The energy-dependent cross sections for $e^+e^- \to \Upsilon(nS)\pi^+ \pi^-$ ($n=1,2,3$) processes
normalized to the leading-order $e^+e^- \to \mu^+\mu^-$ cross sections.
The results of the fits are shown as smooth curves.
The vertical dashed line indicates the energy at which the hadronic cross section is maximal.}
\label{fig:xsec-fits}
\end{figure}

In order to extract the resonance shape, we perform a $\chi^2$
fit to the measured production cross sections, including the one estimated at
$\sqrt{s} = 10.867$ GeV, using the model
\begin{equation*}
{\sigma_{\Upsilon\pi\pi}\over\sigma^0_{\mu\mu}} \propto
A_{\Upsilon\pi\pi} \left|R_0 + {e^{i\phi} BW(\mu, \Gamma)}
\right|^2~, 
\end{equation*}
where $\sigma^0_{\mu\mu} = 4\pi\alpha^2/3s$ is the
leading-order $e^+e^-\to\mu^+\mu^-$ cross section and
$BW(\mu,\Gamma)$ is the Breit-Wigner function
$1/[(s-\mu^2)+i\mu\Gamma]$. 
The normalizations for
$\Upsilon(1S)\pi^+\pi^-$, $\Upsilon(2S)\pi^+\pi^-$, and
$\Upsilon(3S)\pi^+\pi^-$ ($A_{\Upsilon\pi\pi}$), as well as
the amplitude of the flat component $R_0$, the mean $\mu$, width
$\Gamma$, and the complex phase $\phi$ of the parent resonance are
free parameters in the fit. 
Because of the low statistics, common shape parameters ($\mu$, $\Gamma$, $\phi$, and $R_0$) are 
introduced for the three different final states.
Results of the fits, shown as the smooth curves
in Fig.~\ref{fig:xsec-fits}, are
summarized in Table~\ref{tab:xsec-fits}.
The fit quality is $\chi^2 = 24.6$ for 14 degrees of freedom (corresponding to a confidence level of 3.9\%).
An alternative fit without the last data point collected at
$\sqrt{s}\sim11.02$ GeV yields a similar result, $\mu =
10889.0^{+5.8}_{-2.9}$ MeV/$c^2$, $\Gamma = 37_{-10}^{+16}$
MeV/$c^2$, and $\chi^2 = 21.3$ for 11 degrees of freedom (corresponding to a confidence level of 3.0\%).
Systematic uncertainties associated with the cross section
measurements are propagated to the resonance shapes. The fits are
repeated, and the variations on the shape parameters are included
as the systematic uncertainties.
In addition to the uncertainties on the cross sections, 
the beam energy around the $\Upsilon(10860)$ is measured by
$M_{\Upsilon(nS)} + \Delta M$ in the $\Upsilon(nS)\pi^+\pi^-$ events, and an uncertainty of
$\pm1$ MeV (comprising the uncertainties of $M_{\Upsilon(nS)}$ given in Ref.~\cite{ref:PDG2008} and of $\Delta M$ in Ref.~\cite{Abe:2007tk}) is included. For the scan data, a common energy shift is also obtained from the fit to
$\Upsilon(nS)\pi^+\pi^-$ events. The relative beam energies are further checked using the $M(\mu^+\mu^-)$ distributions
of $\mu$-pair samples.

\begin{table}[b!]
\renewcommand{\arraystretch}{1.2}
\caption{
Cross sections at peak ($\sigma_{\rm peak}$), mean ($\mu$), width ($\Gamma$), phase ($\phi$), and the amplitude for the constant component ($R_0$)
from the fit to the energy-dependent
$e^+e^- \to \Upsilon(1S)\pi^+\pi^-$, $\Upsilon(2S)\pi^+\pi^-$, and $\Upsilon(3S)\pi^+\pi^-$
cross sections. Normalization of the resonance term in the fitting model is represented by $\sigma_{\rm peak}$.
There are two solutions for $\phi$ and $R_0$ with identical $\chi^2$.
The first uncertainty is statistical, and the second is systematic.} \label{tab:xsec-fits}
\begin{center}
\begin{tabular}{lc|c}
\hline
\hline
\multicolumn{2}{l}{$\Upsilon(1S)\pi^+\pi^-$ $\sigma_{\rm peak}$ (pb) }  & $2.78_{-0.34}^{+0.42}\pm0.23$ \\
\multicolumn{2}{l}{$\Upsilon(2S)\pi^+\pi^-$ $\sigma_{\rm peak}$ (pb) }  & $4.82_{-0.62}^{+0.77}\pm0.66$ \\
\multicolumn{2}{l}{$\Upsilon(3S)\pi^+\pi^-$ $\sigma_{\rm peak}$ (pb) }  & $1.71_{-0.31}^{+0.35}\pm0.24$ \\
\hline
\multicolumn{2}{l}{ $\mu$ (MeV/$c^2$) }  &  $10888.4_{-2.6}^{+2.7}\pm1.2$ \\
\multicolumn{2}{l}{ $\Gamma$ (MeV/$c^2$) } &  $30.7_{-7.0}^{+8.3}\pm3.1$ \\
& Solution I & Solution II \\
$\phi$ (rad)   &  $1.97\pm0.26\pm0.06$ & $-1.74\pm0.11\pm0.02$ \\
$R_0$ ($($GeV$)^{-2}$)    &  $1.98_{-0.60}^{+0.72}\pm0.20$ & $0.87_{-0.22}^{+0.29}\pm0.09$ \\
\hline
\hline
\end{tabular}
\end{center}
\end{table}

\begin{figure}[t!]
\begin{center}
\includegraphics[width=8.5cm,height=2.915cm]{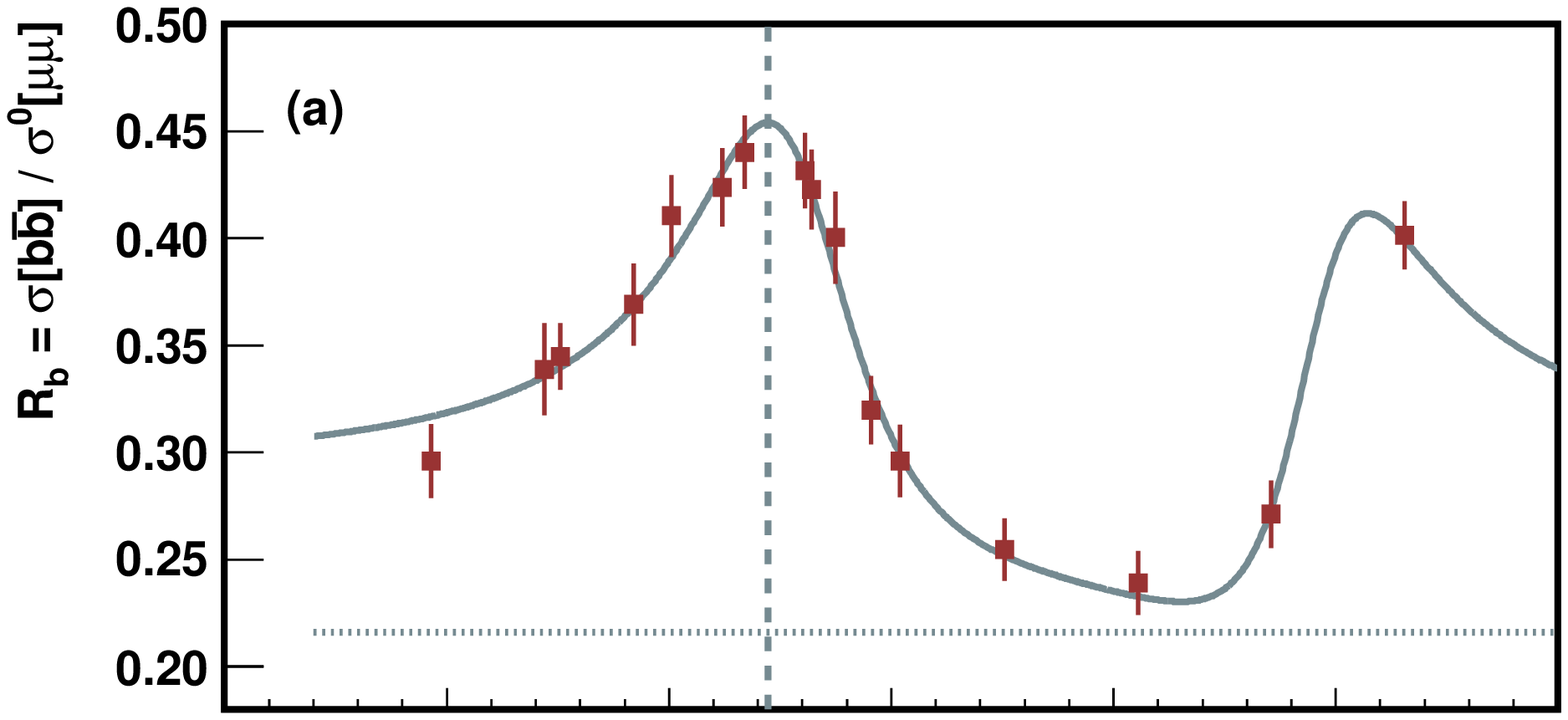}
\includegraphics[width=8.5cm,height=2.915cm]{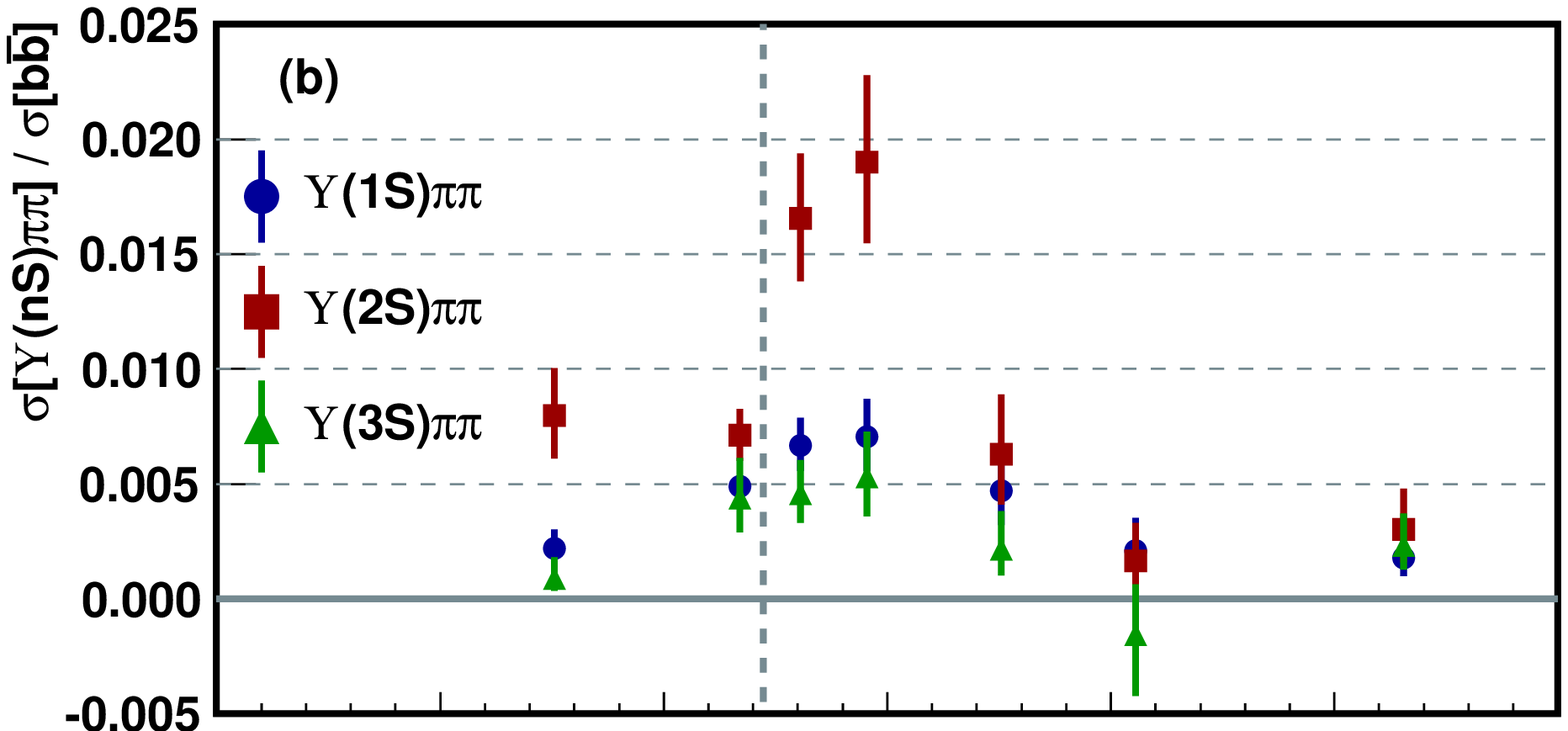}
\includegraphics[width=8.5cm,height=3.5cm]{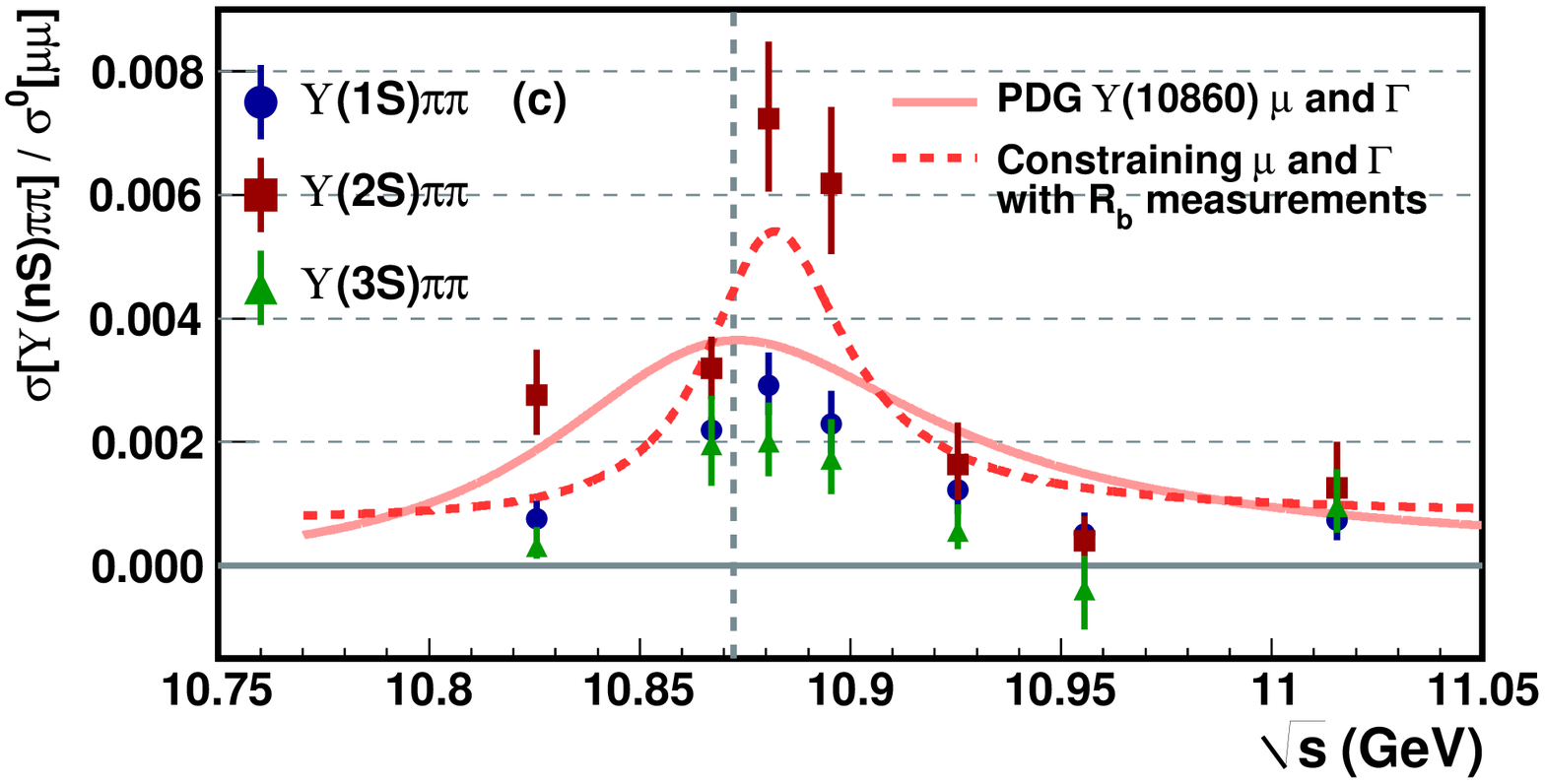}
\end{center}
\caption{(a) $R_b$ and (b)
the ratio between $\sigma[e^+e^-\to\Upsilon(nS)\pi^+ \pi^-]$ and $\sigma[e^+e^-\to b\overline{b}]$ as a function of $\sqrt{s}$;
(c) the energy-dependent cross section ratios for $e^+e^- \to \Upsilon(nS)\pi^+ \pi^-$ events,
the result of fits with resonant parameters from $R_b$ or PDG averages are superimposed.
The horizontal dotted line in (a) is the non-interfering $|A_{nr}|^2$ contribution in the fit.
The vertical dashed line indicates the energy at which the hadronic cross section is maximal.}
\label{fig:had_ratio}
\end{figure}

We also determine the resonance parameters for the $\Upsilon(10860)$ using 
$b\overline{b}$ hadronic events from the energy scan at 
$\sqrt{s}$ between 10.80 and 11.02 GeV.
We measure the fraction $R_b = \sigma_b/\sigma^0_{\mu\mu}$, where $\sigma_b = N_{\rm b}^{R_2<0.2}(s)/\mathcal{L}\epsilon_b(s)$ is
the $e^+e^- \to b\overline{b}$ hadronic cross section. 
The number of $e^+e^- \to b\overline{b}$ events with $R_2 < 0.2$ ($N_{\rm b}^{R_2<0.2}$) is estimated
by subtraction of non-$b\overline{b}$ events scaled from a data set collected at $\sqrt{s} \simeq 10.52$ GeV,
where $R_2$ denotes the ratio of the second to zeroth Fox-Wolfram moments~\cite{ref:R2}.
Selection criteria for hadronic events are described in Ref.~\cite{Abe:2000yh}.
The acceptance for $e^+e^- \to b\overline{b}$ ($\epsilon_b(s)$) is found to vary slightly from 68.1\% to 70.5\% over the range of scan energies.
The line shape used to model our data is given by
$|A_{nr}|^2 + |A_0 + A_{10860}e^{i\phi_{10860}}BW(\mu_{10860},\Gamma_{10860}) + A_{11020}e^{i\phi_{11020}}BW(\mu_{11020},\Gamma_{11020})|^2$;
 this parameterization is the same as that used in Ref.~\cite{ref:BaBar_scan}.
We perform a $\chi^2$ fit to our $R_b$ measurements as shown in Fig.~\ref{fig:had_ratio}(a).
The shapes for $\Upsilon(11020)$ ($\phi_{11020}$, $\mu_{11020}$ , and $\Gamma_{11020}$) are fixed
to the values in Ref.~\cite{ref:BaBar_scan}, since our data points are not able to constrain the $\Upsilon(11020)$ parameters.
The resulting shape parameters for the $\Upsilon(10860)$ are
$\phi_{10860} = 2.33^{+0.26}_{-0.24}$ rad, $\mu_{10860} = 10879 \pm 3$ MeV/$c^2$ , and $\Gamma_{10860} = 46^{+9}_{-7}$ MeV/$c^2$.
These values are consistent with those obtained in Ref.~\cite{ref:BaBar_scan}.
The quality of the fit is $\chi^2 = 4.4$ for 9 degrees of freedom (corresponding to a confidence level of 88\%).

Figure~\ref{fig:had_ratio}(b) shows
the ratio between $\sigma[e^+e^-\to\Upsilon(nS)\pi^+ \pi^-]$ and $\sigma[e^+e^-\to b\overline{b}]$ as a function of $\sqrt{s}$.
As shown in Fig.~\ref{fig:had_ratio}(c), 
an alternative fit applied to the $\Upsilon(nS)\pi^+\pi^-$ cross sections and $R_b$ measurements, simultaneously, with common values of $\mu_{\Upsilon\pi\pi} = \mu_{10860}$ and
$\Gamma_{\Upsilon\pi\pi} = \Gamma_{10860}$ increases the $\chi^2$ by 8.71
and reduces the degrees of freedom by two.
This corresponds to a deviation of 2.5$\sigma$ from the nominal fit
with separated $\mu$ and $\Gamma$ for $\Upsilon(nS)\pi^+ \pi^-$ and $R_b$
measurements.
We also examine the systematic sources as described above, 
and the smallest deviation obtained is 2.0$\sigma$.
If the resonant mean and width are fixed to the results in Ref.~\cite{ref:BaBar_scan} or PDG values~\cite{ref:PDG2008},
deviations of 3.9$\sigma$ or 5.6$\sigma$, respectively, are obtained. 
Scans within the region with
$(\mu_{\Upsilon\pi\pi}-\mu_{10860})^2/\sigma(\mu_{10860})^2 + (\Gamma_{\Upsilon\pi\pi}-\Gamma_{10860})^2/\sigma(\Gamma_{10860})^2 \le 1$,
yield minimum deviations of 3.4$\sigma$ or 5.1$\sigma$, respectively.

In summary, we report the observation of enhanced $e^+e^- \to
\Upsilon(1S)\pi^+\pi^-$, $\Upsilon(2S)\pi^+\pi^-$, and
$\Upsilon(3S)\pi^+\pi^-$ production at $\sqrt{s}$ between
$\sqrt{s}\simeq 10.83$ and $11.02$ GeV. The energy-dependent cross
sections for $e^+e^- \to \Upsilon(nS)\pi^+\pi^-$ events are
measured for the first time, and are found to differ from the
shape of the $e^+e^- \to  b\overline{b}$ cross section. A
Breit-Wigner resonance shape fit yields a peak mass of
$10888.4^{+2.7}_{-2.6} \,({\rm stat}) \pm1.2 \,({\rm syst})$
MeV/$c^2$ and a width of $30.7_{-7.0}^{+8.3} \,({\rm stat}) \pm3.1
\,({\rm syst})$ MeV/$c^2$. A fit excluding the $\sqrt{s}\sim11.02$
GeV data point is consistent with the nominal fit, indicating no
strong contribution of $\Upsilon(nS)\pi^+\pi^-$ events from
$\Upsilon(11020)$. The $\Upsilon(10860)$ shape parameters obtained
from our $R_b$ hadronic cross section 
measurements are consistent with the measurements
from BaBar. The differences between the shape parameters from
$\Upsilon(nS)\pi^+\pi^-$ events and from our $R_b$ measurements are
$\mu_{\Upsilon\pi\pi}-\mu_{10860} = 9 \pm 4$ MeV/$c^2$ and
$\Gamma_{\Upsilon\pi\pi}-\Gamma_{10860} = -15^{+11}_{-12}$
MeV/$c^2$. 
A fit to the $e^+e^- \to \Upsilon(nS)\pi^+\pi^-$ cross sections with our measured $\Upsilon(10860)$ mean and width yields a statistical deviation of $2.5\sigma$ ($2.0\sigma$ including systematic uncertainties).
The $\Upsilon(nS)\pi^+\pi^-$ partial widths were found to be
much larger than the expectations for conventional $\Upsilon(5S)$ states. 
As an extension, energy-dependent $\Upsilon(nS)\pi^+\pi^-$ production 
cross sections are measured; the observed structure 
deviates slightly from the $\Upsilon(10860)$ shape obtained 
from the hadronic cross sections. More data is required to clarify 
the resonance structure.


%
%
%
%


We thank the KEKB group for excellent operation of the
accelerator, the KEK cryogenics group for efficient solenoid
operations, and the KEK computer group and
the NII for valuable computing and SINET3 network support.  
We acknowledge support from MEXT, JSPS and Nagoya's TLPRC (Japan);
ARC and DIISR (Australia); NSFC (China); MSMT (Czechia);
DST (India); MEST, NRF, NSDC of KISTI, and WCU (Korea); MNiSW (Poland); 
MES and RFAAE (Russia); ARRS (Slovenia); SNSF (Switzerland); 
NSC and MOE (Taiwan); and DOE (USA).

\end{document}